\documentclass[3p, 11pt]{core/elsarticle}
\usepackage[acronym]{glossaries}
\makeglossaries

\newacronym{test}{TEST}{This section is not numbered. A nomenclature section could be provided when there are mathematical symbols in your paper. Superscripts and subscripts must be listed separately. Nomenclature definitions should not appear again in the text.
}

\newglossaryentry{test_symbol}
{
        name=$\theta_{test}$, 
        description={This section is not numbered. Define acronyms and abbreviations that are not standard in this section. Such acronyms and abbreviations that are unavoidable in the abstract must be defined at their first mention there. Ensure consistency of abbreviations throughout the article. Always use the full title followed by the acronym (abbreviation) to be used, e.g., reusable suborbital launch vehicle (RSLV), International Space Station (ISS).}
}

\usepackage{makeidx}
\usepackage{amsmath}
\usepackage{epsfig}
\usepackage{amssymb}
\usepackage{textcomp}
\usepackage{pstricks,pst-text,psfrag}
\usepackage{datetime}
\usepackage{setspace}
\usepackage{watermark}
\usepackage{array}
\usepackage{csquotes}
\usepackage{graphicx}
\usepackage{booktabs}
\usepackage{tabularx}

\usepackage{multirow}
\newcommand\Tstrut{\rule{0pt}{2.6ex}}       
\newcommand\Bstrut{\rule[-1.6ex]{0pt}{0pt}} 
\newcommand{\TBstrut}{\Tstrut\Bstrut} 
\usepackage[hidelinks]{hyperref}
\usepackage{cleveref}
\usepackage{longtable}
\usepackage{lscape}
\usepackage{float}
\usepackage{caption}
\usepackage{subcaption}
\usepackage{lscape}

\usepackage{supertabular}

\newcommand{\authorList}    {
Raj Thilak Rajan,
Shoshana Ben-Maor,
Shaziana Kaderali,
Calum Turner,
Mohammed Milhim,
Catrina Melograna,
Dawn Haken,
Gary Paul,
Vedant,
Sreekumar V,
Johannes Weppler,
Yosephine Gumulya,
Riccardo Bunt,
Asia Bulgarini,
Maurice Marnat,
Kadri Bussov,
Frederick Pringle,
Jusha Ma,
Rushanka Amrutkar,
Miguel Coto,
Jiang He,
Zijian Shi,
Shahd Hayder,
Dina Saad Fayez Jaber,
Junchao Zuo,
Mohammad Alsukour,
Cécile Renaud,
Matthew Christie,
Neta Engad,
Yu Lian,
Jie Wen,
Ruth McAvinia,
Andrew Simon-Butler,
Anh Nguyen,
Jacob Cohen
}

\newcommand{\theTitle}
{Applications And Potentials Of Intelligent Swarms \\ For Magnetospheric Studies}

\newcommand{\theAbstract}   {Earth's magnetosphere is vital for today’s technologically dependent society. To date, numerous design studies have been conducted and over a dozen science missions have flown to study the magnetosphere. However, a majority of these solutions relied on large monolithic satellites, which limited the spatial resolution of these investigations, as did the technological limitations of the past. To counter these limitations, we propose the use of a satellite swarm carrying numerous and distributed payloads for magnetospheric measurements. Our mission is named APIS –- Applications and Potentials of Intelligent Swarms. \par

The APIS mission aims to characterize fundamental plasma processes in the Earth's magnetosphere and measure the effect of the solar wind on our magnetosphere. We propose a swarm of 40 CubeSats in two highly-elliptical orbits around the Earth, which perform radio tomography in the magnetotail at 8--12 Earth Radii ($R_E$) downstream, and the subsolar magnetosphere at 8--12 $R_E$ upstream. These maps will be made at both low-resolutions (at 0.5 $R_E$, 5 seconds cadence) and high-resolutions (at 0.025 $R_E$, 2 seconds cadence). In addition, in-situ measurements of the magnetic and electric fields,  plasma density will be performed by on-board instruments. \par

In this article, we present an outline of previous missions and designs for magnetospheric studies, along with the science drivers and motivation for the APIS mission. Furthermore, preliminary design results are included to show the feasibility of such a mission. The science requirements drive the APIS mission design, the mission operation and the system requirements. In addition to the various science payloads,  critical subsystems of the satellites are investigated e.g., navigation, communication, processing and power systems. Our preliminary investigation on the mass, power and link budgets indicate that the mission could be realized using Commercial Off-the-Shelf (COTS) technologies and with homogeneous CubeSats, each with a 12U form factor. We summarize our findings, along with the potential next steps to strengthen our design study.

}

\begin{document}

\title{\theTitle}
\author{\authorList}
\journal{Elsevier Acta Astronautica}
\begin{abstract}
    \theAbstract
\end{abstract}

\maketitle

\newpage
\vspace{5mm}
\section{Introduction} \label{sec:intro} 
\begin{figure*}
    \centering
    \includegraphics[width=0.9\textwidth]{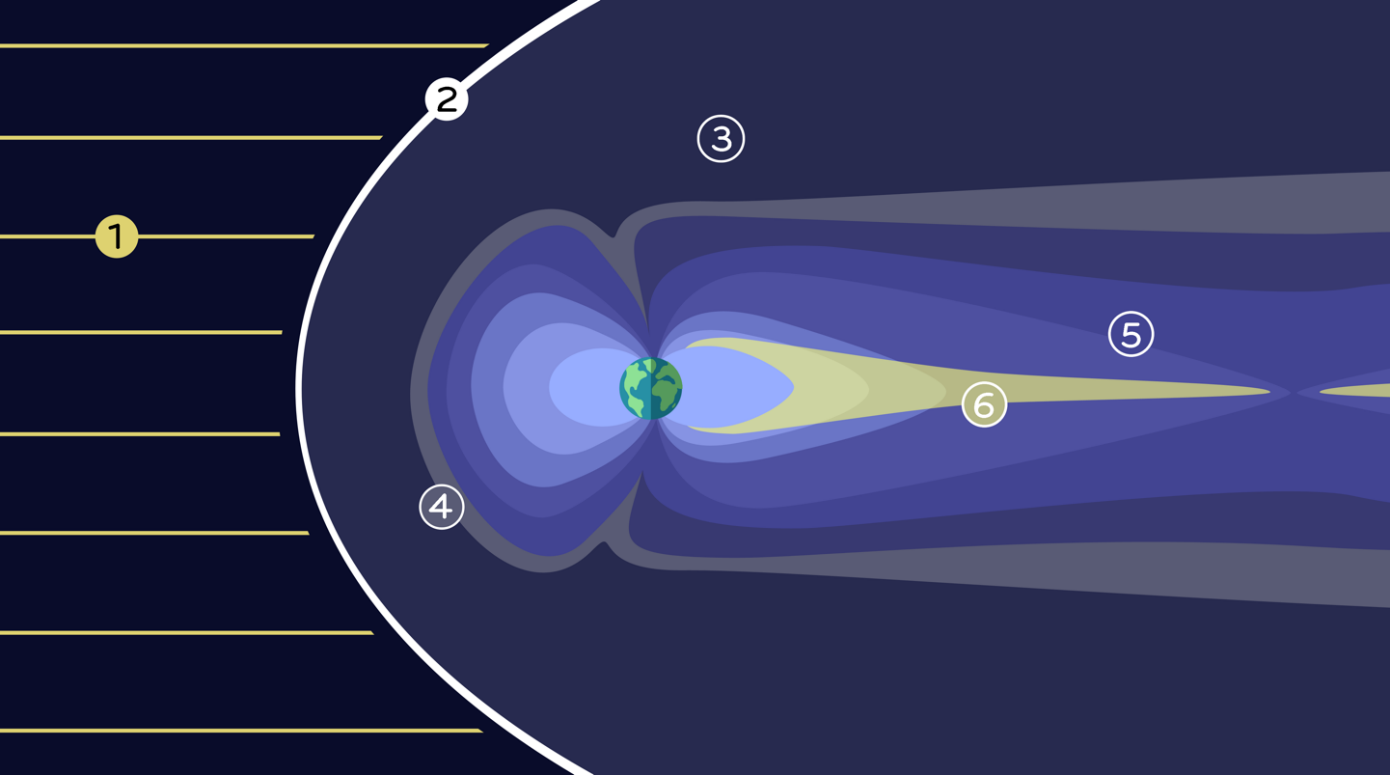}
    \caption{ \textbf{The structure of Earth’s magnetosphere:} Earth’s magnetosphere is shaped by the pressure of the solar wind (1) --- a torrent of charged particles that our Sun ejects outwards to interplanetary space. The solar wind interacts with Earth’s internally generated magnetic field, which decelerates at the bow shock (2), forming a shock wave. The boundary region at which the pressure of the solar wind is equivalent to Earth’s compressed upstream magnetic field is called the magnetopause (4). This region is nearly impenetrable and is located between geosynchronous orbit and the orbit of the Moon. The boundary layer between the plasma bow shock and the magnetopause is the magnetosheath (3),  a transitional region where the density of particles significantly reduces compared to the bow shock. The complex internal structure of the magnetosphere evolves depending on factors such as solar activity. The open magnetic field lines connect to Earth’s polar caps where energetic electrons or protons contribute to aurorae. The magnetic field lines carried by the solar wind sweep in to the magnetotail (5), the teardrop shaped tail of the illustrated magnetosphere. Within the magnetotail, a dense plasma sheet (6) separates the magnetotail’s North and South lobes near the equatorial plane. The APIS mission would investigate both the magnetotail and the Sun-ward magnetosphere, with a particular focus on radio tomography in the magnetotail (5).}
    \label{fig2:MagnetosphereStruct}
\end{figure*}

The heliosphere refers to the area of space under the direct influence of the Sun, which extends from the stellar surface to the outer edges of the solar system. As a space-faring species, we can physically explore this region with satellites, and thereby experimentally test our understanding of heliophysics \cite{schrijver_heliophysics:_2009}. This places heliophysics in a privileged position, as only few other branches of astrophysics can lend themselves to in-situ experimentation, and no other branches impact day-to-day life so profoundly. Heliophysics governs the processes occurring around the stars strewn throughout the universe, and we are in the fortunate position of having a natural laboratory to study these processes close to home. The dominant force in the heliosphere is the solar wind --- the fast-moving, hot and tenuous stream of charged particles constantly emanating from the Sun \cite{kivelson1995introduction}. When the solar wind encounters a planetary magnetic field, it flows around the obstacle like water around a rock \cite{schrijver_heliophysics:_2009}. Astronomers have observed the collision and interaction of the solar wind and planetary magnetic fields across the solar system, from Mercury’s weak magnetic field to the impressive aurorae on Jupiter and Saturn \cite{Smith1979b,Clarke2009,Winslow2013a}. Closer to home, the interaction between the solar wind and Earth’s magnetic field sculpts a structure known as the magnetosphere, within which Earth’s magnetic field is the dominant force \cite{schrijver_heliophysics:_2009}. \Cref{fig2:MagnetosphereStruct} illustrates this concept and shows the structure of Earth's magnetosphere.

\begin{figure*}
    \centering
    \includegraphics[width=.95\textwidth]{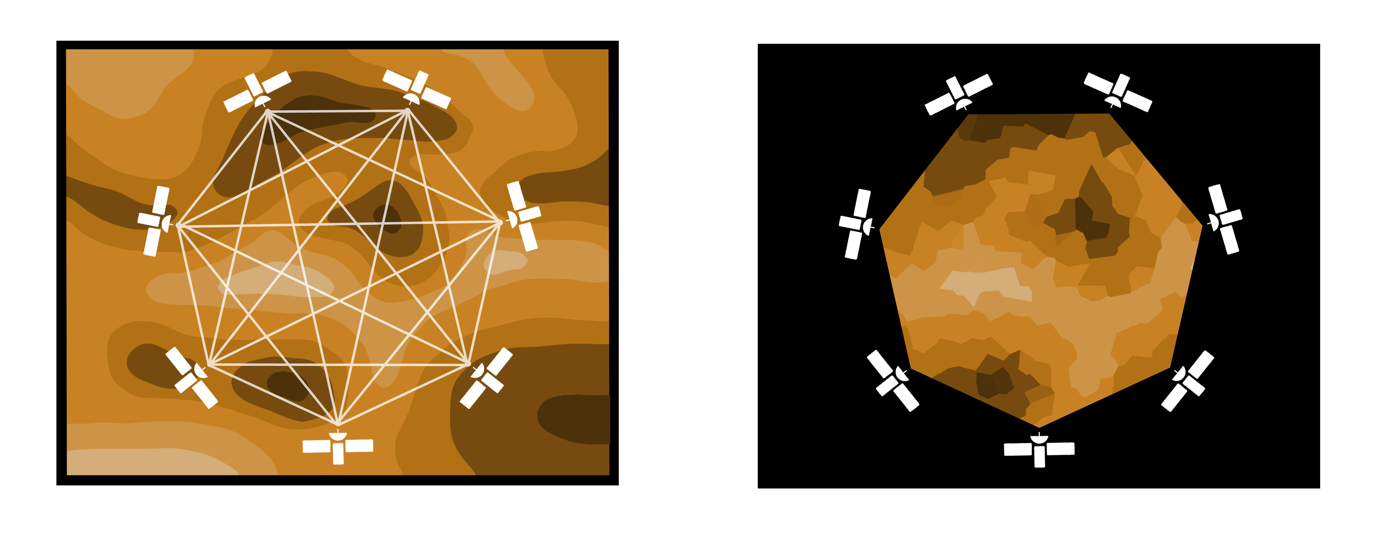}
    \caption{\textbf{An illustration of tomography}: Tomography is the process of imaging using penetrating waves, which are transmitted along the line of sight between nodes through the region of interest. In this illustration, the nodes are satellites and the waves are under consideration are in the radio spectrum. On the left, the image shows the spatial distribution of the plasma and the radio tomography links between the satellites. On the right, the image shows the tomographic reconstruction of the density field within the measured region is illustrated.}
    \label{fig:TomoSketch}
\end{figure*}

\subsection{Motivation} Despite occurring on our astrophysical doorstep, some key physical processes connecting the solar wind and the magnetosphere remain poorly understood \cite{schrijver_heliophysics:_2009},  motivating a steady stream of ongoing research. Understanding the magnetosphere is not only scientifically interesting, but also vital for today’s technologically dependent society. Energy transferred from the solar wind to the magnetosphere triggers electromagnetic storms on Earth, knocking out power grids and infrastructure such as communication networks, navigation, and transport. The effects of these geomagnetic storms on spacecraft can be disastrous \cite{pelton_handbook_2015}. Note, a better understanding and more detailed modeling can help us prepare and possibly prevent any catastrophic situations directly impacting human life. Given these practical and scientific motivations, understanding magnetospheric interactions and processes has been a driving requirement for decades of space science missions. There has been a plethora of studies and missions in the past to measure and study a variety of processes in our magnetosphere. However, despite the numerous heliophysics missions that have flown, plasma turbulence and the formation of plasma structures are still elusive and can only be understood with large-scale multi-point measurements. To achieve this goal, we propose the APIS mission (Applications and Potentials of Intelligent Swarms for magnetospheric studies) --- a swarm of 40 autonomous CubeSats in two highly-elliptical orbits around the Earth, which perform radio tomography in the magnetotail at 8--12 Earth Radii ($R_E$) downstream, and the subsolar magnetosphere at 8--12 $R_E$ upstream. These maps will be made at both low-resolutions (at 0.5 $R_E$, 5 seconds cadence) and high-resolutions (at 0.025 $R_E$, 2 seconds cadence).

\subsection{Outline} We begin our study with an overview of previous missions and case studies in \Cref{sec:previous_studies}, followed by listing the key science drivers for the APIS mission concept in \Cref{sec:science}. The detailed mission design and mission operations are presented in \Cref{sec:mission_design} and \Cref{sec:mission_operations} respectively. The APIS satellites will comprise various science payloads, which are described in \Cref{sec:payloads}. Navigation, communication, on-board processing and power are critical aspects of the APIS swarm, which are addressed in \Cref{sec:navigation}, \Cref{sec:communication},  \Cref{sec:OBC} and \Cref{sec:power_system} respectively. We present the APIS satellite configuration,  summarize our findings, and present directions towards future work in \Cref{sec:conclusion}.





\section{Overview of missions and studies}
\label{sec:previous_studies}

\subsection{Previous missions} In the past decades, more than 20 science missions have flown with the aim of investigating Earth’s magnetosphere, and many more mission proposals exist on paper. After decades of single-point measurements, multi-point imaging of the magnetosphere is pivotal to the upcoming heliophysics missions. These missions are more realizable given the rise in efficient small satellite architectures designs \cite{barnhart2007very}, the growing market in cubesat commercial vendors \cite{behrens2019exploring,buchen2015small} and a demand for various Earth observation missions \cite{sandau2010status}. In general, the space industry is witnessing a major disruption guided by innovation and determination to take bold risks \cite{reddy2018spacex}.

One notable mission that has gathered distributed in-situ measurements is NASA’s ongoing mission, THEMIS \cite{Angelopoulos2008b}. Launched in 2007, the mission originally comprised of five satellites (each weighing 134 kg) in the magnetotail, and has provided over 12 years of data collection to date. The payload addresses the science goal of investigating substorms --- magnetic phenomena that release energy and intensify aurorae. Another mission capable of collecting three-dimensional information on Earth's magnetic environment and its interaction with the solar wind is ESA’s Cluster mission, which was launched in 2000 and is still operational. The four Cluster satellites flying in a tetrahedral formation probe the interactions of electrons and waves in Earth's magnetic environment \cite{Escoubet2001}.

A relevant upcoming CubeSat mission is NASA’s CubeSat for Solar Particles (CuSP), which is designed to study solar particles in a three-month mission and to act as a pathfinder for a network of “Space Weather Stations”. The mission is planned to launch with the Space Launch System in 2021 and consists of a single $30\times20\times10$ cm CubeSat in a Sun-pointing, trans-Lunar, heliocentric orbit at 1 AU. On a larger scale, NASA’s Magnetospheric Multiscale (MMS) mission provides unprecedented high‐time resolution multi-point particle and field measurements \cite{Karlsson2018b}. Launched in 2015, MMS has a highly eccentric orbit and operates in Earth's magnetosphere using four identical spacecraft flying in a tetrahedral formation. MMS researches the microphysics of magnetic reconnection, energetic particle acceleration, and turbulence -- processes that occur in astrophysical plasmas. Additionally, a group of micro-satellites performed simultaneous multi-point measurements of Earth’s magnetic field as part of NASA’s ST5 (Space Technology 5). This  90-day mission mission flew in 2006 and tested 10 new technologies to pave the way for future multi-satellite missions such as MMS and THEMIS. Furthermore, ST5 contributed to an early understanding of the magnetosphere’s dynamic nature \cite{Young2007a}. In recent years, the growing capabilities of small satellites have also led to a variety of innovative distributed heliophysics missions being proposed, developed, and funded \cite{green2020scientific,plice2019helioswarm}. These missions leverage low launch costs and off-the-shelf hardware to offer scientific returns at a lower cost than traditional monolithic satellites \cite{millan2019small}. However, in comparison to APIS, these missions typically comprise of either a single CubeSat or a handful of small satellites in a constellation \cite{blum2020gtosat,bishop2019low,klenzing2020petitsat,deforest2019polarimeter,westerhoff2015laice,hecht2019daili,Desai2019a}.

\subsection{Case studies} \label{subsec:CaseS} The science community has recognized for decades the need for satellites to make simultaneous, distributed heliophysics measurements. The distributed architecture of swarm satellites lends itself to high spatio-temporal range measurements \cite{Klein2019a}, making it a promising architecture for such distributed measurement missions. Two case studies of proposed magnetic constellations provide insights on how the APIS swarm will perform a next generation heliophysics mission. The case study missions are NASA’s MagCon and MagCat, both of which were designed to probe Earth’s plasmasheet and magnetotail. NASA planned these missions to determine how the magnetosphere stores, processes, and releases energy in the magnetotail and accelerates particles to the inner radiation belts. The secondary scientific objective of the mission was to study how Earth’s magnetosphere responds to variable solar wind and how this influences the magnetopause, the boundary between the solar wind and Earth’s magnetic field. However, neither mission proceeded further than a concept due to budget constraints. 

\textit{The Magnetospheric Constellation (MagCon)}: The proposed Magnetospheric Constellation mission was designed to perform distributed in-situ measurements of the magnetic field, plasma, and particles in such a way as to \textit{“revolutionize our understanding of the magnetospheric response to dynamic solar wind input and the linkages across systems”} \cite{Guana}. The mission concept was a constellation of up to 36 small satellites weighing 30 kg each with a typical spacing of 1–2 $R_E$ (Earth Radii), using orbits with perigees in the 7--8 $R_E$ range and apogees dispersed uniformly up to 25 $R_E$ \cite{national_research_council_solar_2013}. Each spacecraft was designed with a boom-mounted magnetometer and a three-dimensional plasma analyzer to measure Earth’s magnetic field. A simple energetic ion-electron particle telescope was also included to analyse charged particle energization, loss, and transport throughout the heliosphere.

\textit{The Magnetospheric Constellation and Tomography (MagCat)}: The Magnetospheric Constellation and Tomography mission was designed to provide the first global images of the magnetosphere \cite{NationalAcademyofSciences2016a}. The mission was designed to examine plasma plumes in the magnetosphere, acquire reconstructed images of plasma density and turbulence using radio tomography, and measure three-dimensional ion and electron distributions. Tomography, that create maps of plasma density \cite{Zhai2011b}, is a key facet of the APIS mission.

\section{Science drivers for APIS} \label{sec:science} The aforementioned missions and studies have significantly improved our understanding of small-scale physical processes in the magnetosphere, such as magnetic reconnection and plasma currents \cite{Angelopoulos2008}.  Despite the numerous heliophysics missions that have flown, plasma turbulence and the formation of plasma structures are still elusive and can only be understood with large-scale multi-point measurements \cite{national_research_council_solar_2013,Matthaeus2019}. In 2004, NASA stated that MagCon’s database of dispersed measurements would allow us to \textit{“emerge from a long and frustrating hiatus”} \cite{NASAScienceandTechnologyDefinitionTeamfortheMagnetosphericConstellationMission2004b}.

\subsection{Motivation} The APIS mission would use a suite of instruments to bridge this gap by providing large-scale maps of plasma density and turbulence in the magnetotail — a need that was identified as early as 2000 \cite{Ergun2000}. The swarm architecture will allow the temporal and spatial resolution of the tomographic maps to vary over the mission. By providing these high-resolution maps, the APIS mission will address two of the four key science goals set out in the 2013 decadal survey on heliophysics \cite{national_research_council_solar_2013}. The baseline tomography measurements meet the spatial resolution (0.5 $R_E$), and cadence (15 seconds) targets set out by the decadal survey, and the high-resolution operational modes comfortably exceed both of these targets. Based on the decadal survey and past, current, and future heliophysics missions, the APIS mission has two main scientific goals:

\begin{itemize}
\item \textbf{Goal 1: Discover and Characterize Fundamental Plasma Processes in the Magnetosphere.}
The APIS mission shall measure the plasma flows and turbulence in the magnetotail using radio tomography and in-situ measurements. The use of a swarm architecture to produce high-resolution, small-scale tomographic maps, as well as large-scale observations will help explain key plasma processes that occur, not only in the magnetosphere, but also in magnetized plasmas across the universe \cite{schrijver_heliophysics:_2009}. These processes, such as turbulence in a magnetized plasma, require multi-scale multi-point measurements to be fully understood \cite{Matthaeus2019}.

\item \textbf{Goal 2: Determine the Dynamics and Coupling of Earth’s Magnetosphere and the Response to Solar Inputs.} The operational architecture of the APIS mission shall allow simultaneous plasma density measurements of both the magnetotail and the Sun-facing magnetosphere. The data provided by the APIS mission will uncover relationships between plasma density between different parts of the magnetosphere. Our orbital design would allow for detailed multi-plane measurements of plasma density in the magnetotail, providing long-awaited data to the heliophysics community at unprecedented spatial and temporal resolutions \cite{Ergun2000}. This data will provide an insight into the plasma dynamics of the magnetosphere in response to solar variation.
\end{itemize}
\subsection{Regions of interest} The APIS mission will provide measurements in two initial regions of interest by launching two groups of swarm satellites. The two regions of interests will be in, (a) A near-equatorial orbit in the magnetotail, and (b) A polar orbit that sweeps through the magnetotail and the Sun-ward magnetosphere, over the course of one year. The region of interest for both equatorial and polar obits is 8–12 $R_E$ from Earth, where a host of scientifically interesting processes occur. After creating large-scale tomographic maps, the satellite swarm will then move on to the second phase of science operations and produce high-resolution maps of selected areas within the magnetotail.

\textit{The Magnetotail}: One group of satellite swarms will be placed in a highly eccentric polar orbit, travelling through the magnetotail --- the teardrop shaped tail of the magnetosphere streaming away from the Sun shown in \Cref{fig2:MagnetosphereStruct}. Our region of interest is $8$--$12$ $R_E$, where key physical processes such as magnetospheric instabilities, plasma flows, morphological changes associated with geomagnetic storms, and turbulence occur \cite{Ergun2000,Angelopoulos2008,Matthaeus2019}. Thus, this region has been intensely studied by previous missions, although at smaller spatial scales than we propose. Initial science observations will provide the big-picture data required to understand the region and the processes happening in this region. The high-resolution follow-up observations will then study these processes in more detail.

\textit{Polar plane}: A secondary plane of the APIS mission swarm satellites will orbit on the same scale ($8$--$12$ $R_E$), nearly perpendicular to the first plane. This plane will sweep through the magnetotail and the Sunward magnetosphere over the course of a year. While in the magnetotail, the group of satellites in the polar plane will be able to enhance downstream measurements by increasing the region of focus. While in the sub-solar magnetosphere, the swarm satellites will be able to measure Sun-side dynamics. The measurements of plasma densities both up- and down-stream of Earth will uncover couplings and dynamics in Earth’s magnetosphere.

\subsection{Science measurements} It is currently not possible to directly image the large-scale plasma structure in the magnetotail, as in-situ measurements require an extremely high number of satellites to achieve the desired resolution. Therefore, we propose to use radio tomography to reconstruct the spatial distribution of plasma. Tomography is the process of imaging with the use of penetrating radio waves. The radio waves are transmitted along many intersecting lines of sight through a region of interest. The density integral along each of these lines is derived from the delay in transmission. With many of these line integrals, an estimate of the density map of the region can be produced, as shown in \Cref{fig:TomoSketch}. Tomographic methods are well-developed for medical imaging, with examples including Computed Tomography (CT) scanning and medical ultrasound. Thus, the APIS mission will use radio tomography to create estimates of the plasma density in the magnetotail. Radio signals will be transmitted between the satellites and the time delay of each signal will be measured. The time delay of the signals is directly related to the total plasma density along the line-of-sight, and a map of the plasma density can then be reconstructed by mathematically combining the line-of-sight density measurements.

The study of the magnetosphere using spacecraft has been proposed \cite{Ergun2000} and tested \cite{Zhai2011b}. Outside the magnetosphere, International Sun-Earth Explorers 1 and 2 have demonstrated the ability to derive the electron density in the solar wind through radio wave propagation \cite{Etcheto1978a}. With measurements between multiple spacecraft, the APIS mission will be capable of investigating the large-scale plasma density structure.


\subsection{Science requirements} The primary region of interest for the mission is the magnetotail, in the range of 8--12 $R_E$ from Earth, and the desired resolution of the tomographic reconstruction is 0.5 $R_E$. Given the desired resolution of the resultant tomographic image $R_S$, and the effective diameter of the area of interest $d$, the approximate number of linear integrations required is: $N>\pi d/R_S$. The orbital characteristics and the number of spacecraft must meet this driving science requirement.

The radio tomography used by APIS relies on the propagation delay of radio waves within the plasma. The satellites of the APIS mission would use two steps to measure the characteristic delay along the line-of-sight.  The first step is the differential phase measurement \cite{Leitinger1994}, which requires two coherent radio signals with different frequencies. The phase velocity of the radio signal in the plasma depends on the frequency and the plasma density. The phase of a probing frequency is compared to the phase of a reference frequency transmitted through the plasma, and the resulting phase delay depends on the plasma density as follows: \begin{equation}
    \Delta \phi_1 = \left( \frac{\omega_1 e^2}{2 \varepsilon_0 m_e c} \right) \left( \frac{1}{\omega_1^2}  - \frac{1} {\omega_{ref}^2} \right) \int n dL
    \label{eq:TomoPhase}
\end{equation} where, $\omega_1$ is the probing frequency, $e$ is the electron charge, $\varepsilon_0$ is the vacuum permittivity, $m_e$ is the electron mass, $\omega_{ref}$ is the reference frequency, and $n$ is the total electron content \cite{Ergun2000}. If the phase delay is greater than $2\pi$, then the second step of delay measurement must be employed. The group delay  resolves the phase delay that is proportional to total electron content: \begin{equation}
    \Delta t_g  \cong - \frac{e^2}{2 \varepsilon_0 m_e c}  \left( \frac{1}{\omega_1^2}  - \frac{1} {\omega_{ref}^2} \right) \int n dL
    \label{eq:TomoPhase_group}
\end{equation} where $\int n dL$ is directly proportional to the total electron content \cite{Davies1989}. By combining the methods of differential phase delay (\ref{eq:TomoPhase}) and group delay (\ref{eq:TomoPhase_group}), we can derive the total electron content. A long-wavelength probing frequency is desired because both types of delay are inversely proportional to frequency. In-situ measurements of the magnetic field and the plasma density are required to interpret the tomography measurements correctly. The scientific measurements must also be correlated with time and position data to produce a complete picture of the magnetotail environment. \Cref{tab:SciRequirements,tab:SciAccuracyRequirements} encapsulate the scientific motivation and requirements of the APIS mission, as well as the measurement accuracies required. These requirements are based on the decadal survey and previously proposed missions, which can be feasibly achieved with current technology \cite{Ergun2000,Angelopoulos2008,Desai2019a,national_research_council_solar_2013}.

\begin{table*}
\scriptsize
\caption{Science requirements for the APIS mission, where Extended, Baseline and Threshold denote the NASA terminologies for best case, baseline and minimum requirements respectively.}
\footnotesize{
    \begin{tabular}{|p{4cm}|| p{3.5cm} |p{3cm} |p{3cm}|}
         \hline
         \textbf{Requirement} & \textbf{Extended} & \textbf{Baseline} & \textbf{Threshold} \Tstrut\\ [0.5ex] 
         \hline\hline
         \textbf{Large-Scale Radio \newline Tomography Phase} & \footnotesize{Multi-plane tomographic maps of the magnetotail 8--12 $R_E$ downstream, and the subsolar magnetosphere at 8--12 $R_E$ upstream with 0.5 $R_E$ resolution. All measurements at 5 seconds cadence.} & \footnotesize{Tomographic maps of the magnetotail 8--12 $R_E$ downstream, and the subsolar magnetosphere at 8--12 $R_E$ upstream with 0.5 $R_E$ resolution. All measurements at 10 seconds cadence.} & \footnotesize{Tomographic maps of the magnetotail 8-12 $R_E$ downstream with 0.5 $R_E$ resolution. All measurements at 15 seconds cadence.} \Tstrut\\ [1ex]\hline
        \textbf{Fine-Scale Radio \newline Tomography Phase} & \footnotesize{High-resolution tomographic maps of small regions of the magnetotail 8--12 $R_E$ downstream at 0.025 $R_E$ spatial resolution. Measurements at 2 seconds cadence for short bursts.} & \footnotesize{High-resolution tomographic maps of small regions of the magnetotail 8--12 $R_E$ downstream at 0.05 $R_E$ spatial resolution. Measurements at 3 seconds cadence for short bursts.} & \footnotesize{High-resolution tomographic maps of small regions of the magnetotail 8--12 $R_E$ downstream at 0.1 $R_E$ spatial resolution. Measurements at 5 seconds cadence for short bursts.} \Tstrut\\[1ex]\hline
         \textbf{In-situ measurements to anchor tomography} & \footnotesize{Measurements of magnetic field, electric field, plasma energy distribution, plasma density; at 2 seconds cadence.} & \footnotesize{Measurements of magnetic field, plasma energy distribution, and density; at 3 seconds cadence.} & \footnotesize{Measurements of magnetic field and plasma density at 5 seconds cadence.} \Tstrut\\ \hline
        \textbf{Positional Knowledge} & 0.01 $R_E$ (63.71 km)& 0.01 $R_E$ (63.71 km) & 0.01 $R_E$ (63.71 km) \Tstrut\\ [1ex]\hline
         \textbf{Time Knowledge} & 0.01 microseconds & 0.01 microseconds & 0.01 microseconds \Tstrut\\ [1ex]\hline    
         \textbf{Duration of Science \newline Observations} & 1 Solar cycle (11 years) & 2 years & 4 months \Tstrut\\ [1ex] 
         \hline
    \end{tabular}}
\label{tab:SciRequirements}
\end{table*}

\begin{table*}
\centering
\caption{Science measurement requirements of the APIS mission (all measurements will have a variable cadence of up to 2 seconds)}
\footnotesize{
    \begin{tabular}{|p{6.2cm}|| p{3cm} |p{3.5cm} |} 
         \hline
         \textbf{Measurement} & \textbf{Range} & \textbf{Resolution} \Tstrut\\ [0.5ex] 
         \hline\hline
        \textbf{Radio Tomography Plasma Density} & 0.05--150 cm$^3$ & 2 \% error acceptable \Tstrut\\ [1ex] \hline
         \textbf{Magnetic Field} & 0--2000 nT & 0.025 nT  \Tstrut\\ [1ex] \hline
         \textbf{Plasma Particle Energy} & 1–5 MeV & 15-20\%   \Tstrut\\ [1ex]\hline
        \textbf{In-situ plasma density} & 0.05--150 cm$^3$ & 2\% error acceptable \Tstrut\\ [1ex] 
         \hline
     \end{tabular}}
\label{tab:SciAccuracyRequirements}
\end{table*}


\subsection{The APIS mission} \label{sec:mission_concept} The APIS mission will address key physical processes in the magnetosphere, including: how plasma enters the magnetosphere; the formation and dynamics of the plasma sheet; the formation of plasma structures in response to solar wind variability; and turbulence in a magnetized plasma \cite{national_research_council_solar_2013}. By exploiting the swarm architecture, the APIS mission will provide large-scale, high-resolution tomographic maps that exceed the targets set out in the 2013 decadal survey on heliophysics \cite{national_research_council_solar_2013}. The novel scientific feature of the APIS mission is the swarm-enabled ability to vary the spatial and temporal resolution of the tomography measurements, which will provide the precise data needed to understand key heliophysics processes. The APIS mission requirements - as discussed at length in this article - are summarized in Table~\ref{tab:MissionRequirements}, and are derived from the science requirements as described in \Cref{tab:SciRequirements} and \Cref{tab:SciAccuracyRequirements}.


\begin{table*}
\centering
\caption{APIS mission overview}
\begin{tabular}{|l|l|} 
\hline
\multicolumn{2}{|c|}{\textbf{Mission requirements}}                                                        \\ 
\hline
\textbf{Launch capability}    & 40 spacecrafts in 2 elliptical orbits                                    \\ 
\hline
\textbf{Mission duration}       & 4 months science phase                                                   \\ 
\hline
\textbf{Orbital requirement}    & 8--14 $R_E$, one polar and one near-equatorial plane  \\ 
\hline
\multicolumn{2}{|c|}{\textbf{Spacecraft requirements }}                                                    \\ 
\hline
\textbf{Attitude stabilization} & 3-axis stabilized                                                        \\ 
\hline
\textbf{Mass}                   & 21.8 kilogram~ (see \Cref{tab:MassBudg})                                                       \\ 
\hline
\textbf{Power}                  & 63.2 watts (see \Cref{tab:PowerBudg})                                                          \\ 
\hline
\textbf{ISL Data rate}          & $\ge$ 100 kbit/sec                                      \\ 
\hline
\textbf{Pointing}               & $\leq 5^\circ$ accuracy, $\leq 2^\circ$ knowledge                                                                  \\
\hline
\end{tabular}
\label{tab:MissionRequirements}
\end{table*}

\begin{figure}
    \centering
    \includegraphics[width=0.45\textwidth]{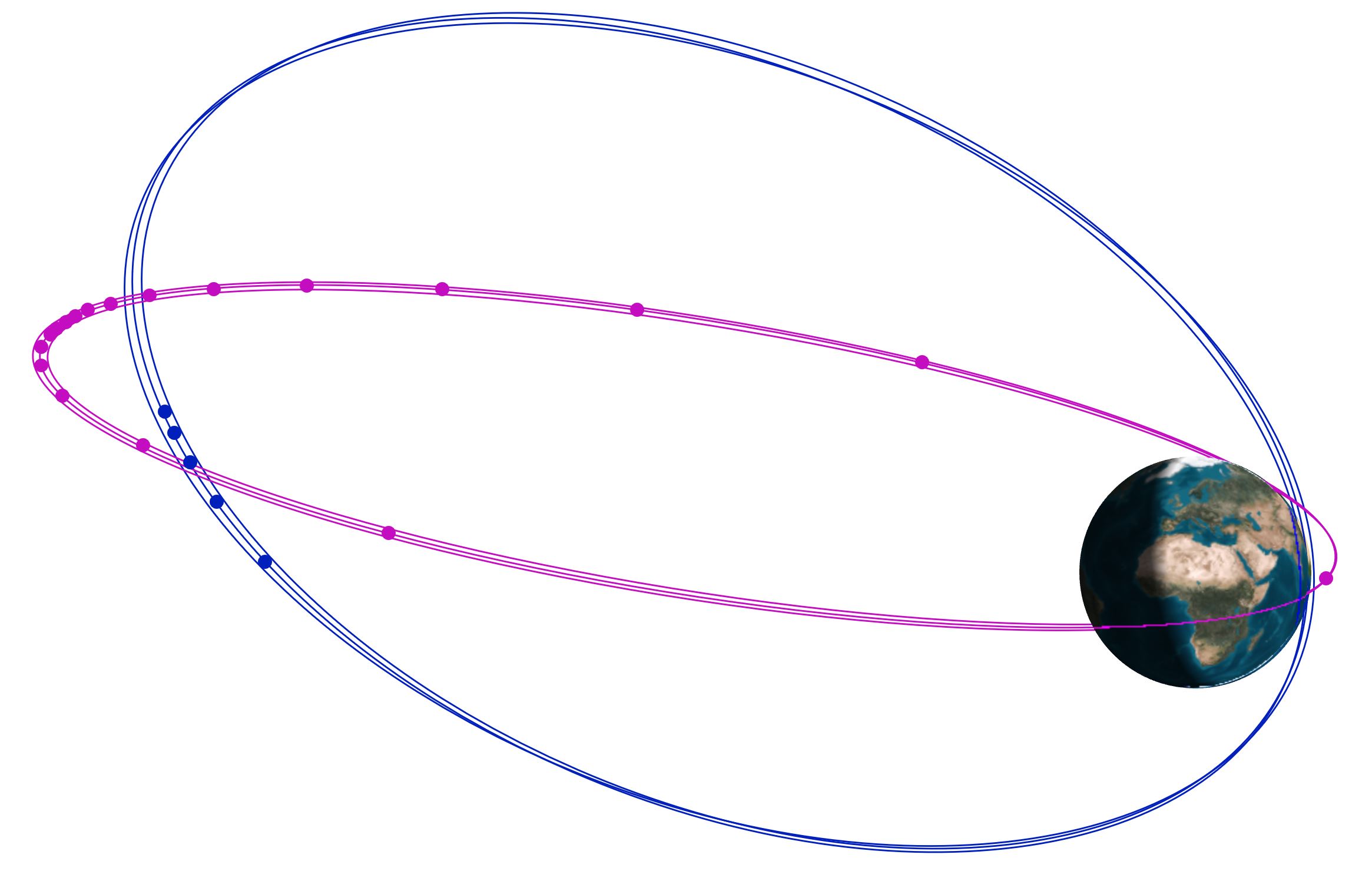}
     \caption{ \textbf{Mission Design:} An illustration of the two orbits in the APIS mission --- i.e., the Polar orbit (in blue), and the Near-Equatorial (in pink)}.
    \label{fig6:MissionDes}
\end{figure}

The APIS mission architecture comprises of a homogeneous satellite swarm spread over two orbits, polar and near-equatorial, as illustrated in \Cref{fig6:MissionDes}, using the Systems Tool Kit (STK). The swarm satellites will exhibit emergent behavior through cooperation in order to achieve tomography measurements, reference-free calibration of instruments, navigation, and data handling. The presented orbits meet a set of requirements, derived from the science objectives. \Cref{tab8:MissOrbParam} shows the mission orbital parameters, where the Right Ascension of the Ascending Node (RAAN) and the argument of perigee are excluded at this time, since the two will depend on the time and date of the launch.


\begin{table*}[tb]
    \centering
    \scriptsize
    \caption{Mission orbital parameters}
    \begin{tabular}{|m{4cm}|c|m{7cm}|}
    \hline
        \textbf{Parameters}     & \textbf{Value} \TBstrut    & \textbf{Comments} \\ \hline\hline
        \textbf{Apogee} &
          13 $R_E$ (altitude) &
          14 $R_E$ from the center of Earth in order to enclose the region of interest, defined in by the science objectives. \TBstrut\\ \hline
        \textbf{Perigee}   & 500 km (altitude) & Higher than ISS orbit \TBstrut\\ \hline
        \textbf{Inclination}   & $\sim 0 ^\circ$ and $\sim 90 ^\circ$   & One inertially locked in polar, other precesses  \\ \hline
        \textbf{Number of satellites in each plane} & 20   & Train formation, 18$^\circ$ of angular separation \Tstrut\\ \hline
        \textbf{Eccentric anomaly \newline of the $\mathbf{n^{th}}$ satellite} & $i \frac{360}{20} n ^\circ $ &
          18$^\circ$ of angular separation, $n$ is the satellite number (between 1 and 20) \\ \hline
    \end{tabular}
    \label{tab8:MissOrbParam}
\end{table*}

\begin{figure*}
     \begin{subfigure}[t]{0.55\textwidth}
         \centering
         \includegraphics[width=\textwidth]{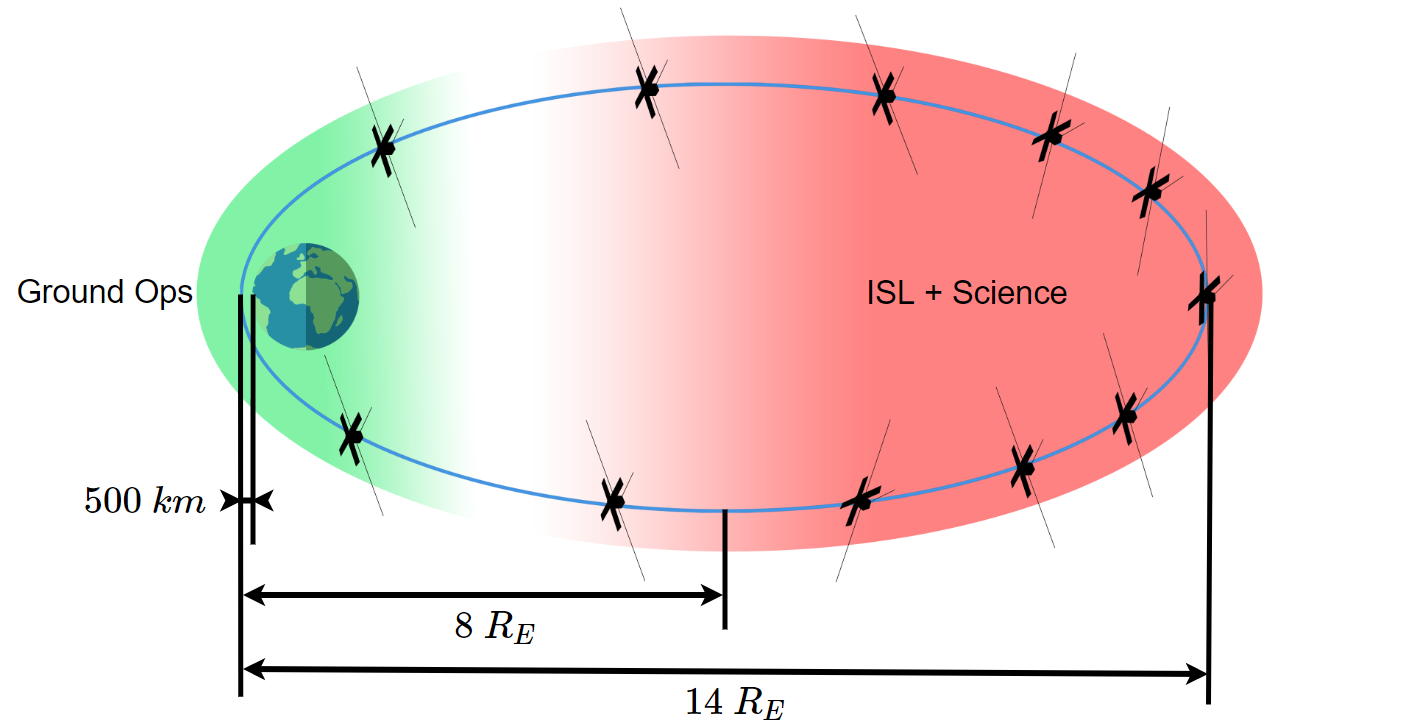}
        \caption{}
         \label{fig7:MissOpPhases}
     \end{subfigure}
     \begin{subfigure}[t]{0.45\textwidth}
         \centering
         \includegraphics[width=0.7\textwidth]{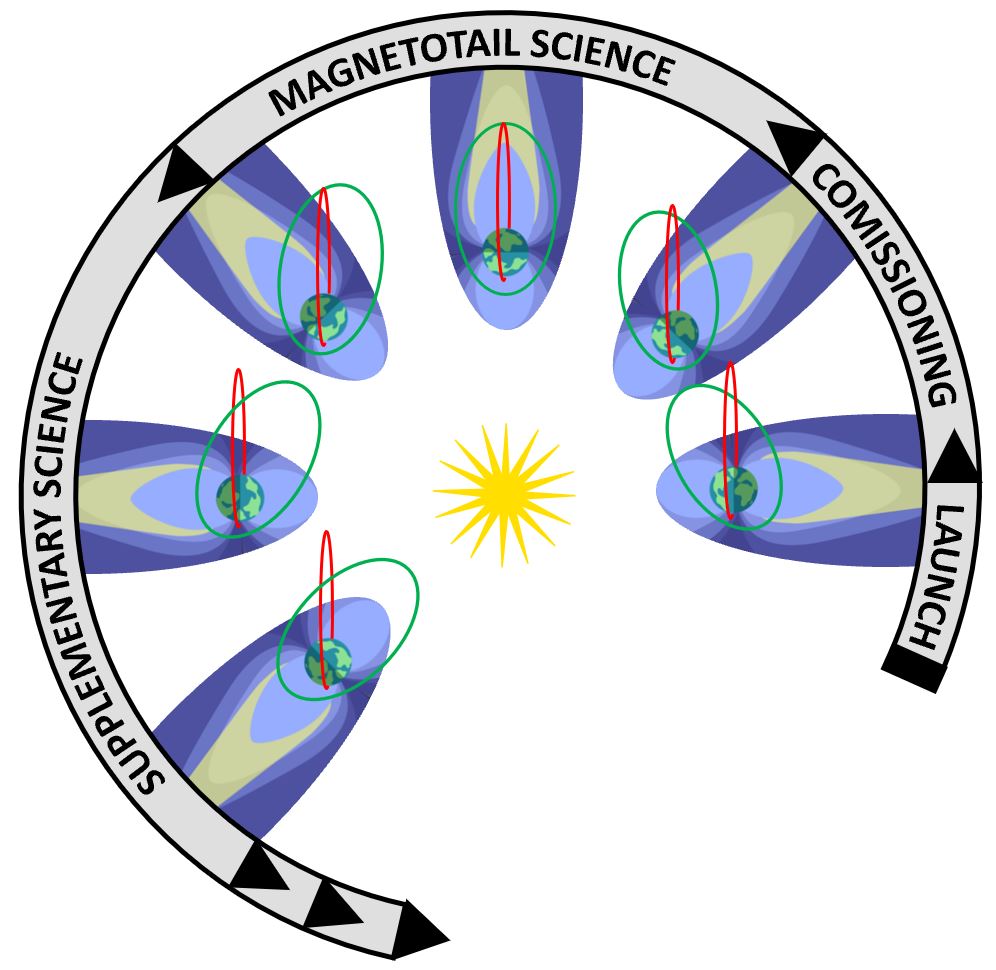}
        \caption{}
         \label{fig13:OrbConfig}
     \end{subfigure}
     \caption{\textbf{Illustration of the APIS mission operational phases and configurations:} (a) The orbital configuration of the swarm over the course of the one-year mission is illustrated (b) The phases of the APIS mission are illustrated, over the course of an orbital period for each satellite in the swarm.  }
    \label{fig:Phases}
\end{figure*}

Over the course of one orbit, the mission operations are subdivided into various phases as illustrated in Figure~\ref{fig7:MissOpPhases}, which include the science phase and the ground operation phase. The green region indicates the ground link for the mission. Due to the highly eccentric orbit, each satellite will have a window of ~$1$-$2$ hours per orbit for communication with the ground station. The satellites will distribute data amongst the swarm and queue the data in order of importance to the science mission for downlink. The sharing and queuing allows for an effective downlink window of up to 20 hours. The region between the perigee and 8 $R_E$ is dedicated for other functions, such as  orbital maneuvering and attitude control. 

The red region in \Cref{fig7:MissOpPhases} indicates the science phase of the orbit, where the swarm performs tomography and plasma measurements. If one of the satellites recognizes an interesting physical phenomena, then the satellite needs to direct the rest of the swarm to take increased measurements through a consensus approach. At the apogee the satellites slow down due to orbital dynamics and the distance between them is reduced to enable better communication between the agents. Therefore, in addition to the science case, the inter-satellite links are also established during this phase, which is critical for swarm-related data processing. Furthermore, individual agents perform science logging preferentially during the ascending phase (agents moving from perigee to apogee), which ensures that inter-satellite data exchange occurs near apogee.


\section{Mission design} 
\label{sec:mission_design}

The objective of the science mission is to map the plasma densities at the scale of $0.5~R_E$, for a region of interest from $8~R_E$ to $12~R_E$. To achieve our goal we have designed the swarm to fly in two orbits perpendicular to one another at inclinations of $90 ^{\circ} $ and approximately $0 ^{\circ}$, with an apogee radius of $14~R_E$, enclosing the magnetospheric area of interest. This configuration enables tomography in the magnetotail, as well as in the Sun-ward magnetosphere. In  Figure~\ref{fig13:OrbConfig} the orbital dynamics that occur during the first year of the mission are illustrated, and the mission phases timeline is described in \Cref{tab14:MP}. 

\begin{table*}[tb]
\centering
\scriptsize{
  \caption{Mission Phases Timeline}
  \label{tab14:MP}
  \resizebox{\textwidth}{!}{%
\begin{tabular}{| m{3.5cm} || m{8cm} | p{1.8cm} | p{1.8cm} |}
\hline
\textbf{Phase} & \textbf{Description} & \textbf{Mission Timeline} & \textbf{Duration}  \Tstrut\\ [0.5ex]\hline\hline
\textbf{Launch} & Launch into highly eccentric orbit & 1st day & $<$1 day \Tstrut \\ 
 \hline
\textbf{Deployment and \newline commissioning} & Swarm agents are deployed at apogee into two possible inclinations (0$^\circ$ or 90$^\circ$), 20 agents per orbit. Subsystems and payloads are validated and calibrated.
 & 1st day & 3 weeks \Tstrut\\
 \hline
 \textbf{Science Phase 1} & Swarm agents cooperate to perform tomography and plasma property measurements in the magnetotail region. & 3rd week   & 3.5 months \Tstrut\\
 \hline
 \textbf{Science Phase 2}  & Cooperative tomography and plasma measurements in the Sun-ward magnetosphere region are performed on the equatorial orbit, and later on the polar orbit. & 6th month & 3 months  \Tstrut\\
 \hline
 \textbf{Science Phase 3}  & Reconfiguration of swarm agents is executed on both orbits to perform higher resolution tomography in a $2~ R_E$ strip of the magnetotail region & 1 year   & 4 months \Tstrut\\
  \hline
\textbf{Maintenance and \newline Ground Operations}  & Desaturation of reaction wheels, charging
downlink of scientific measurements prioritized by swarm agents, and data analysis.
 &\multicolumn{2}{l|}{All mission} \Tstrut\\
   \hline
\textbf{End of life}& Passivation of subsystems, natural orbit decay, comprehensive data analysis  &\multicolumn{2}{l|}{Dependent on orbit decay} \Tstrut\\[1ex]
 \hline 
\end{tabular}}
}
\end{table*} 

\begin{figure*}
     \centering
     \begin{subfigure}[t]{0.4\textwidth}
         \centering
         \includegraphics[width=1.1\textwidth]{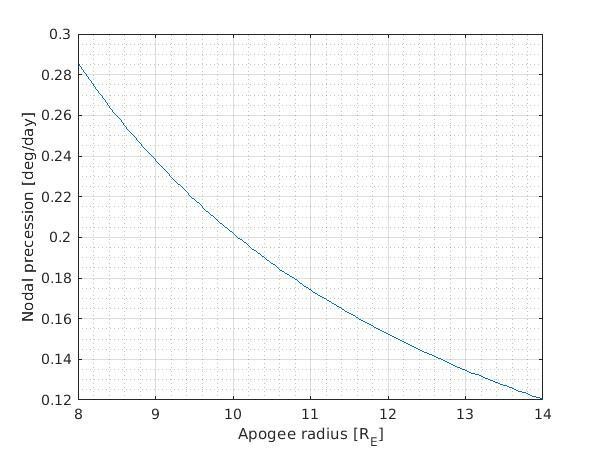}
        \caption{}
         \label{fig14:NodalDecay}
     \end{subfigure}
     \hfill
     \begin{subfigure}[t]{0.5\textwidth}
         \centering
         \includegraphics[width=1.1\textwidth]{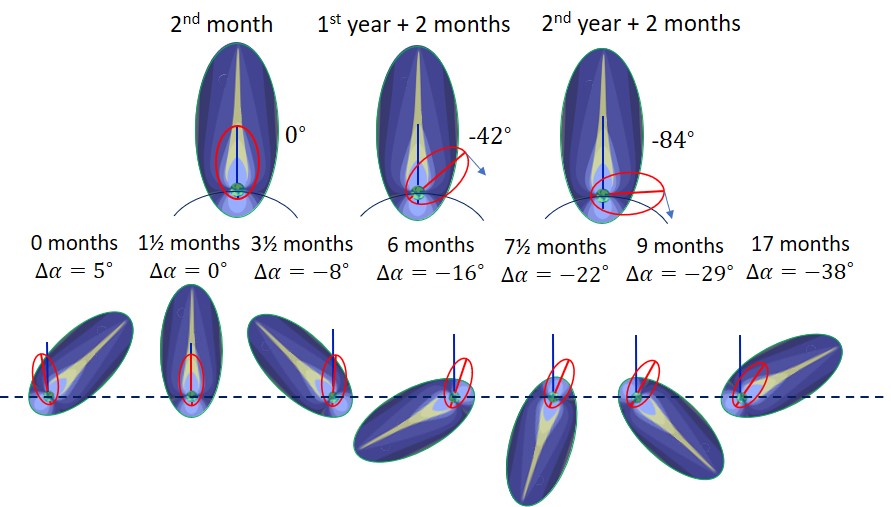}
        \caption{}
         \label{fig15:TailPrecess}
     \end{subfigure}
     \caption{ \textbf{Variation of nodal precession against apogee:} (a) The equatorial orbit used in the mission will experience nodal procession rates of 0.12$^\circ$--0.29$^\circ$ per day while the apogee radius is 8--14 $R_E$. (b) Precession of the equatorial orbit relative to the magnetotail.}
    \label{fig:Phases}
\end{figure*}

\subsection{Nodal precession of the Orbit} \label{subsec:NPO} One of the main features of the mission orbit design is the differing nodal precession that occurs between the equatorial and polar orbits, causing the semi-major axis of both orbits to no longer be aligned. Orbital precession affects the orientation of the elliptic trajectory of spacecraft. Nodal precession is defined as the rotation of the orbital plane around the axis of the central body, Earth in our case. This phenomenon is caused by non-uniform mass distribution in the central body, and in a first approximation, the major contributor is the equatorial bulge of Earth that causes the planet to be an oblate spheroid with a larger diameter at the equator than at the poles. The relationship that relates the precession rate to the orbital elements is the following:
\begin{equation}
    \omega_p = -\frac{3}{2}\frac{R_E^2}{(a(1-e^2))^2}J_2\omega \cos{(i)}
\end{equation}
where, $a, e,\omega$ and $i$ define the orbit, $R_E$ is the equatorial Earth radius, and $J_2$ represents the perturbations due to the oblateness of the Earth. From the two orbits in the APIS mission, only the equatorial orbit is affected by nodal precession. The polar orbit, as characterized by an inclination of 90$^\circ$, nullifies the precession due to the $cos(i)$ term in the equation given. While the satellites pass through the region of interest and the apogee radius remains between $8~ R_E$ and $14~ R_E$, where science measurements are to be performed, the experienced precession rates will range between $0.12 ^\circ$ and $0.29 ^\circ$ per day. \Cref{fig14:NodalDecay} illustrates how the regression rate varies while the orbit decays in relation to the phases, as illustrated in \Cref{fig15:TailPrecess}.

\subsection{Orbital maneuvers} \label{sec:OM} The minimum science requirement is achieved in the course of the first 3--4 months into the mission, with tomography in the magnetotail. The mission then continues for an additional two months to perform measurements of the plasma density outside the tail. Following this timeframe, the extended mission will exploit the swarm ability to reconfigure and take new measurements. The objective of the new geometry is to perform higher resolution tomography, as the satellites sweep through the magnetotail in the next orbital pass about $12$ months into the mission, which is detailed in \Cref{tab14:MP}. 

\begin{figure*}
    \centering
    \includegraphics[width=0.8\textwidth]{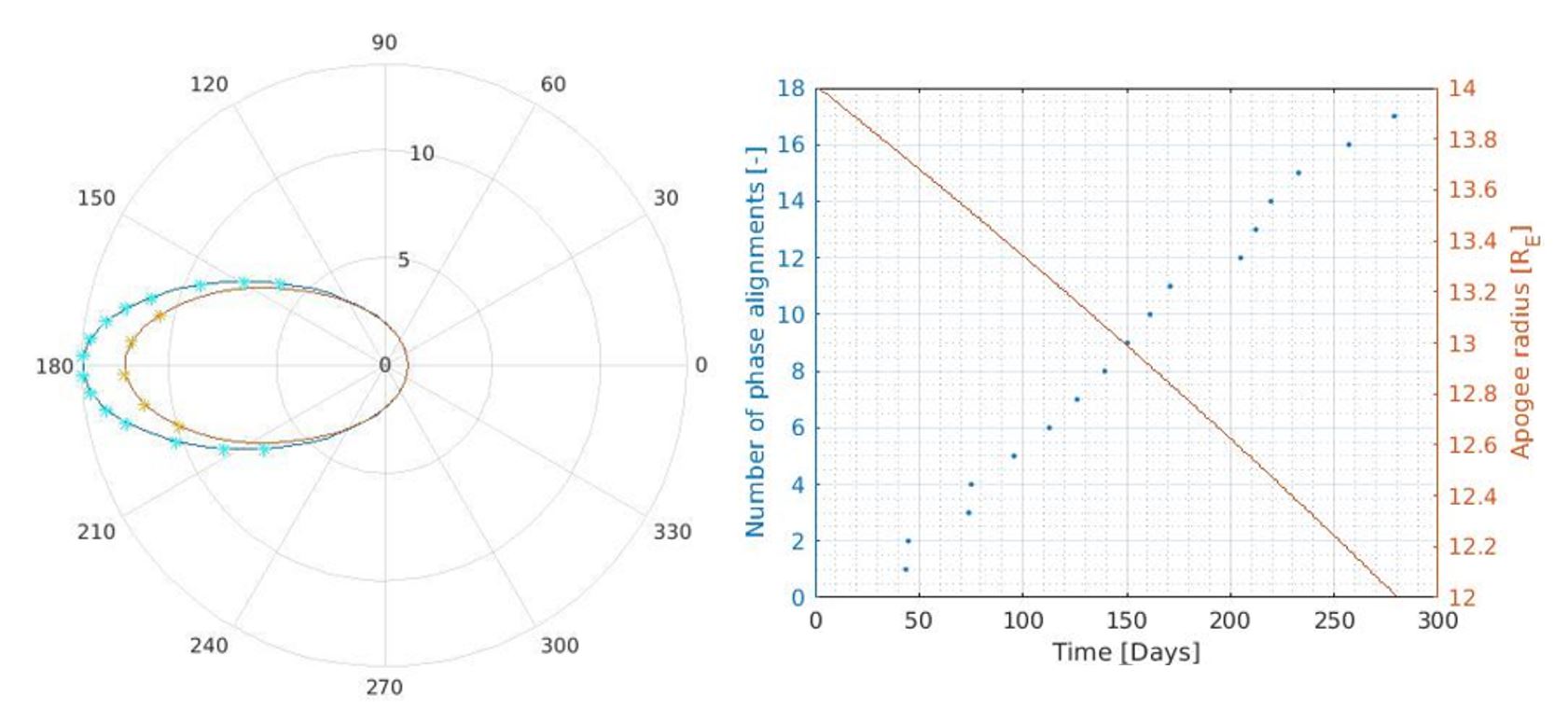}
    \caption{ \textbf{Effect of differing orbital periods between groups 1 and 2:} Frequency of apogee alignment between a fixed reference orbit and a reconfigured high resolution orbit is presented.}
    \label{fig17:ApFreq}
\end{figure*}

\textit{Reconfigurability}: \label{subsub:recon}In the proposed scenario, five satellites from each orbit will perform a maneuver to change their apogee to $2~R_E$  lower than the reference orbit at $14 ~R_E$. The remaining $30$ satellites will stay in their original orbits. The maneuver will take approximately nine months, commencing outside the magentotail and terminating before the satellites sweep through the magnetotail again. Modifying the trajectory of the APIS satellites requires significant thrust due to the high energy of the orbit, as well as a regenerative propulsion system for the extended mission duration. To overcome this difficulty, we will use the electric propulsion system on each APIS satellite, which can operate over long periods of time (see \Cref{sec:Prop}). The atmospheric drag experienced during flight close to perigee can also be utilized. One of the major trade-offs to be considered is the effect of different orbital periods between Group 1 (in the near-equatorial orbit) and Group 2 (in the polar orbit). In \Cref{fig17:ApFreq}, the apogee alignment between the two groups is presented, as the orbit decays and the period changes. The two groups are aligned when they reach apogee within one hour of one another.

In a first approximation, the electric propulsion system will enable an apogee of $30$ km per orbit when used to reduce the velocity in perigee. Attitude control orients the satellites to maximize the satellite surface area exposed to the thin atmosphere. The minimum surface area, used to maintain the $15$ satellites in the reference orbit, is $0.06$ m$^2$. The maximum possible area exposing the solar panels is $0.52$ m$^2$, which induces a drag of more than eight times greater than otherwise. Indicatively, the orbit decay, induced by drag on the five satellites moving to the lower orbit, is $20$ km per orbit at a perigee altitude of $500$ km. A first estimate of the total decay rate induced combining both electric propulsion and drag  is in the order of 50 km per orbit.

The swarm is expected to implement emergent behavior to achieve collaboration between the two orbital groups, thereby improving the potential science return. This emergent behavior arises from simple rules followed by the satellites, without the need for a centralized coordination, and thus enabling the  APIS swarm to collectively behave as a single entity. The desired behavior is to modify the altitudes of both orbits, maximizing the frequency of apogee alignment. This modification will require each satellite to plan the desired attitude at perigee to modulate both the amount of drag experienced and the thrust generated due to turning the propulsion system on and off. 



\subsection{End of mission} As the swarm decays below the region of interest of $8~ R_E$, the APIS mission enters the final stage, and no more research is viable. Rapid decay will help minimize the number of spacecraft in orbit. The orbit apogee will decay to a $500$ km, placing teh satellites into a circular orbit due to the effects of atmospheric drag on the satellites. Once in the circular orbit, the final resources from the propulsion system will be used to lower the perigee and apogee below the altitude of the ISS. Then the satellites will orbit at the maximum drag attitude orientation while passing through perigee. This procedure will allow for rapid de-orbit, reducing possible intersection with the ISS orbit at lower altitudes. 


\section{Mission operations} \label{sec:mission_operations}
Over the course of the full APIS mission, the satellite swarm will undergo key operational phases. In this section we present an overview of the four mission phases: launch, deployment and commissioning, science, and decommissioning.



\subsection{Launch phase} Our proposed orbit design outlines a scenario of 40 spacecraft distributed over two orbital planes. The difference in inclination between the two planes is 90$^\circ$. In addition to mass, an important requirement for the launcher is its fairing volume. The total volume to be occupied by the APIS satellites is 0.48 m$^3$. A single launcher will first inject the whole swarm into a highly eccentric equatorial orbit. Then, once at apogee, a kick-stage will provide enough $\Delta v$ to change the orbital plane for 20 of the 40 satellites from equatorial to polar. Performing the maneuver at apogee will optimize the use of propellant mass. The following equation provides the $\Delta v$ required: \begin{equation}
    \Delta v = 2v_a \sin{ \bigg(\frac{\Delta i}{2}\bigg)}
\end{equation} where $v_a$ is the velocity at apogee, and $i$ is the inclination of the orbital plane. For the APIS mission the final stage will impart a $\Delta v$ of $1.13$ km/s to a payload mass of $520$ kg.  The SpaceX Falcon 9 is a launcher that best meets the requirements of the mission, considering both cost and capabilities. The block 5 iteration of the rocket is capable of positioning 8300 kg in Geostationary Transfer Orbit (GTO) \cite{spaceX_falcon9}. Thus, it would be possible for the APIS satellites to share the launcher with other missions, therefore reducing the cost. Placing the apogee from the geostationary belt to the required orbit will require a further $\Delta v$ of $820$ m/s. 

\subsection{Deployment and Commissioning} All satellites will be attached to an EELV\footnote{Evolved Expendable Launch Vehicle} Secondary Payload Adapter (ESPA) ring, which is an adapter for launching secondary payloads on orbital launch vehicles and has become a \textit{de facto} standard for various spaceflight missions \cite{jeremy_espa_2001,wegner2001eelv,maly2017espa}. The commissioning phase is automated, and each swarm satellite tests their payload and bus systems. The satellites transmit any anomalies to the ground for further investigation. Due to the swarm capability, the swarm group in one orbital plane can start its science phase without the need to wait for the swarm group in the second plane. Once all satellites are deployed and commissioned in both orbital planes, the science phase will generate results at a significantly higher rate.

\subsection{Science phase} 
\begin{figure*}
     \centering
     \begin{subfigure}[t]{0.45\textwidth}
         \centering
         \includegraphics[width=0.7\textwidth]{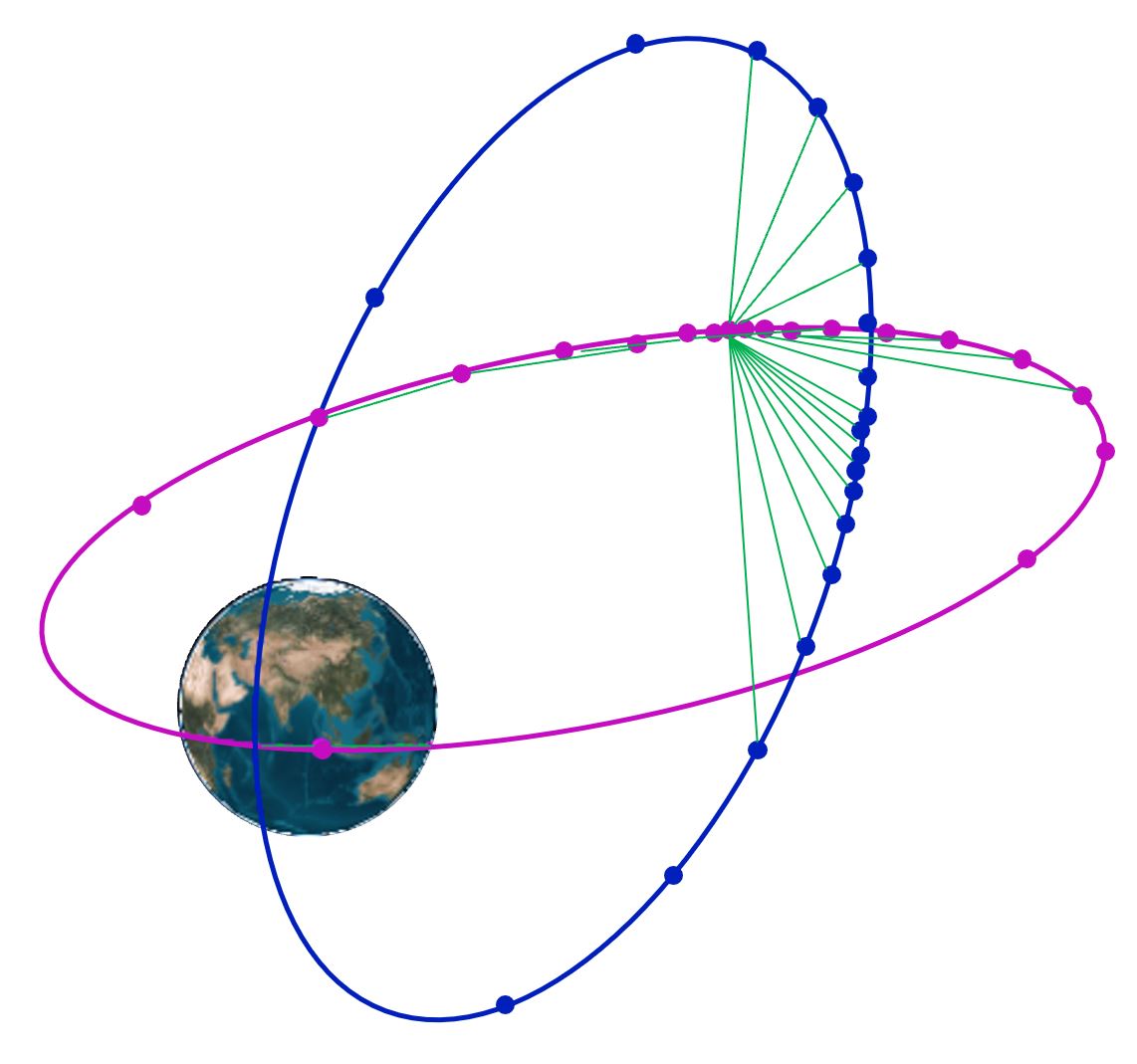}
        \caption{}
         \label{fig19:ISLSim}
     \end{subfigure}
     \hfill
     \begin{subfigure}[t]{0.45\textwidth}
         \centering
         \includegraphics[width=0.7\textwidth]{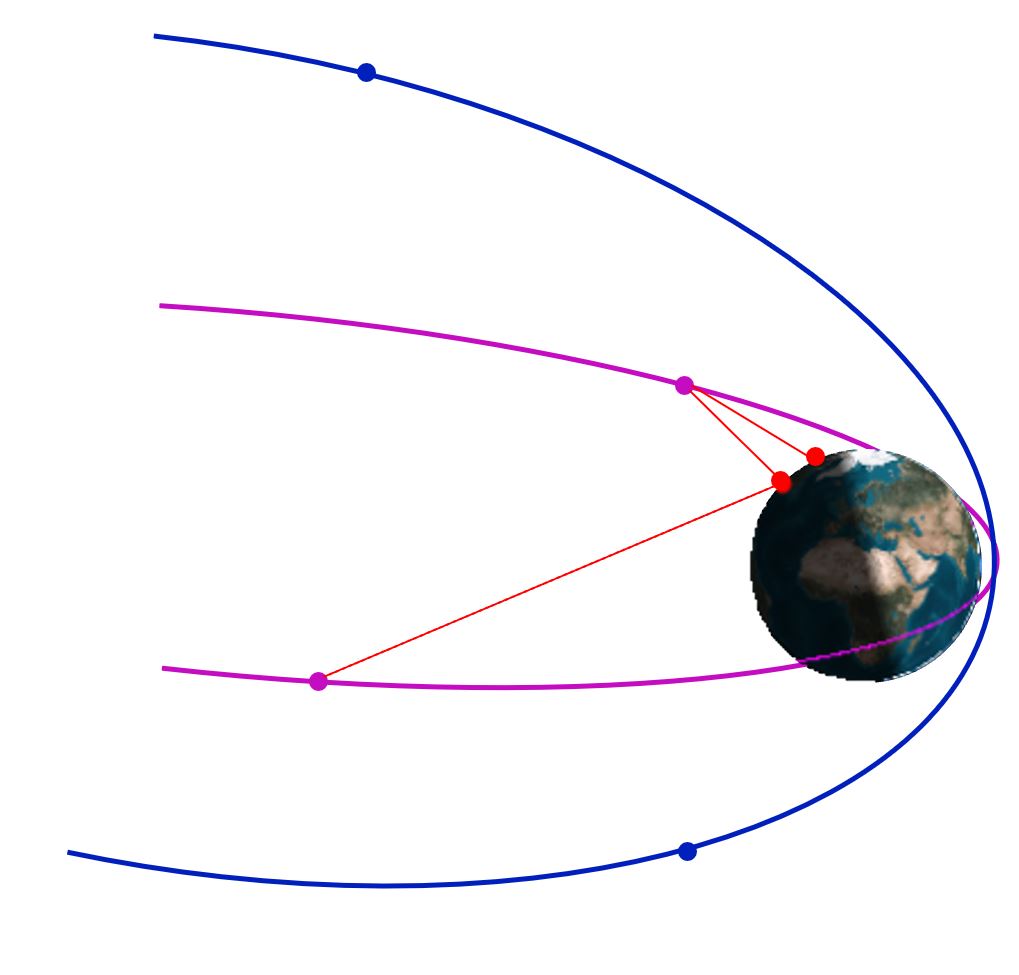}
        \caption{}
         \label{fig12:CommswGrnd}
     \end{subfigure}
     \caption{ \textbf{(a) Inter-Satellite Link (ISL):}  An illustration of APIS ISLs within the swarm in the APIS mission. \textbf{ (b) Downlink:}  Swarm satellites can communicate with ground stations while in the equatorial orbit.  Red lines indicate ground link communication with satellites.}
    \label{fig:Phases}
\end{figure*}

The majority of the mission's lifetime is the science phase. The orbital design allows the swarm satellites to be in the magnetospheric region of interest by use of orbit precession. The science phase (at an altitude above 50000 km) lasts for 27 hours per orbit. Tomography techniques require a minimum number of lines to reach the necessary spatial resolution. The swarm will achieve this minimum threshold by performing tomography measurements when close to the apogee area, where the satellites cluster together at a range of $2~R_E$ to $8 ~R_E$ separation distance due to orbital dynamics.  The transmissions will synchronize, using the Code-Division Multiple Access (CDMA) protocol, which leads to the notable swarm behavior. An illustration of these Inter-satellite Links (ISLs) within the swarm is shown in \Cref{fig19:ISLSim}.



Each satellite will also perform in-situ measurements of the particle distribution and plasma properties for the tomography data processing. During the 4-month science phase, the satellites will perform 1000 measurement sequences. After each satellite stores its data, the swarm will perform autonomous measurement prioritization based on information theory, which maximizes the scientific return of the swarm. While clustered at apogee, ISL communications are enabled between neighbouring satellites for measurement sharing, data sorting, and autonomous decision-making on measurements to be kept in the collective swarm memory for maximizing the scientific data quality.



\section{Science payloads} \label{sec:payloads} All the satellites in the APIS mission have a payload suite comprising of multiple science payloads, including the radio tomography system, and various in-situ measurement systems. In addition, all APIS satellites will employ on-board Radio-Frequency Interference Mitigation (RFIM) techniques and sensor calibration to ensure the veracity of the recorded science data.


\subsection{Radio tomography} \label{sec:radio_tomography} Large-scale in-situ measurements of the entire field of interest would require a significant number of satellites, however, with our proposed design, we use radio tomography in order to estimate the plasma density map. In radio tomography imaging, each satellite transmits a coherently phased pair of discrete radio signals, which in turn is received by all satellites. The measured phase difference between the signals, integrated along the ray path, yields the Total Electron Content (TEC). For a network of 20 satellites, with a maximum distance separation of 8 $R_E$, each inter-satellite transmission and reception takes up to 0.3 seconds, and the entire tomographic cycle takes up to 3 seconds \cite{Ergun2000}. The choice of the frequency pair --- i.e., the probing and reference frequencies --- plays a vital role in the mission design. For example, in the MagCat mission (see \Cref{subsec:CaseS}), the probing and reference frequencies were of 1 MHz and 3 MHz --- i.e., the third harmonic was used \cite{Ergun2000}. To transmit and receive at these selected wavelengths, a half-wave dipole antenna of approximately 50 m is required, which increases both the mass and power budget of the small satellites.

\begin{figure*}[tb]
    \centering
    \includegraphics[width=0.7\textwidth]{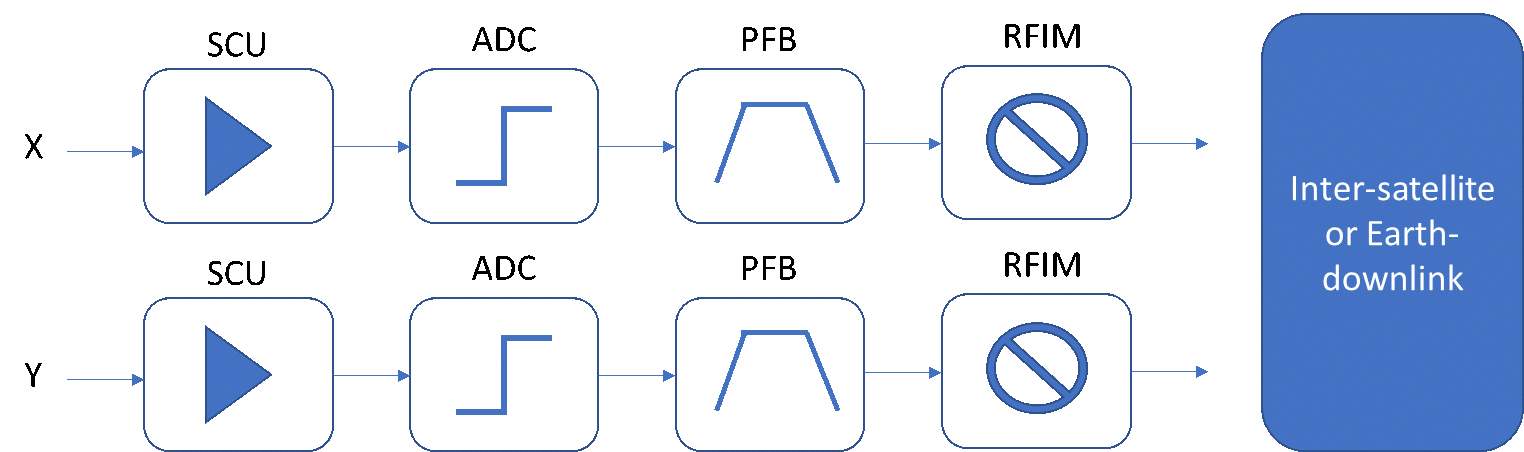}
    \caption{ \textbf{Signal processing for RFI mitigation: } The block diagram shows a breakdown of the signal processing for radio tomography within a single APIS satellite.}
    \label{fig:rifm}
\end{figure*}

To overcome power and mass limitations, the APIS mission plans to use 10 MHz and 30 MHz, as the probing and reference frequencies, respectively. The necessary antenna lengths at these wavelengths is approximately 15 m, which is suitable for the small satellites of the mission. Radio measurements at these selected frequencies can become significantly corrupted by man-made interference \cite{Ergun2000}. Thus, the APIS mission employs the on-board RFIM technique to resolve this issue. The dual polarized radio signals received by each satellite is to be pre-processed by the Signal Conditioning Unit (SCU), digitized by the Analog to Digital Converter  (ADC), and filtered by a Poly-phase Filter Bank (PFB). Finally, the man-made interference is removed by the RFIM block, as illustrated in \Cref{fig:rifm}.

The swarm satellites are to deploy two 7.5 m dipole antennas, which will unfurl in measuring tape fashion. The boom material is a carbon fiber reinforced polymer, while the design is a combination of two arms in the form of ‘C’-sections with one inside the other to reduce volume. The boom will also act as an insulator, and thus, will be wrapped with aluminized Kapton or a similar coating to control electrostatic discharge \cite{Miyazaki2018}.

\subsection{In-situ measurements} \label{sec:in-situ_measurements} In addition to the radio tomography payload, there are $3$ additional payloads on-board the satellites for in-situ measurements.
\begin{itemize}
    \item \textit{Super-thermal Ion Spectrograph}: The super-thermal ion spectrograph is based on the electrostatic analyzer payload proposed for the CuSP mission \cite{Desai2019a}. The payload occupies a volume of $1.5$U. The payload provides a measure of the energy spectra and the peak intensities of the incident particles, thereby providing in-situ measurements of the magnetospheric plasma.
    \item \textit{Miniaturized Electron and Ion Telescope}: This payload is based on a flight-proven payload flown on the Compact Radiation Belt Explorer mission in 2018. Our payload will be shielded with tungsten and aluminum, which increases the mass but is necessary to reduce the background noise caused by scattering. The APIS mission miniaturized electron and ion telescope instrument contains a stack of SSDs to image the electron and ion paths. Similar payloads have been flown on-board the ISS \cite{barao2004ams}.
    \item \textit{Vector Helium Magnetometer}: Vector helium magnetometers have been used in previous missions to measure magnetic fields in various environments. These magnetometers measure the optical properties of helium, which change in an applied magnetic field. Recent and upcoming missions have miniaturized this instrument, and it is now adapted for CubeSat missions like CuSP \cite{Desai2019a}.
\end{itemize}

\subsection{Calibration} Calibration is a key challenge for these on-board detectors, especially in long-term multi-year missions. Conventionally, operators on Earth correct the errors accrued by the detectors over time via telemetry. As an alternative, the swarm satellites can communicate with each other to employ relative or reference-free calibration, to eliminate the detector errors such as offsets and gain \cite{Rajan2018}. Since the signal subspace is unknown, the nodes can employ blind calibration \cite{Balzano2007}, which would involve a training phase during the initial deployment of the antennas. Distributed calibration algorithms will be used to ensure the APIS satellite swarm could collectively alleviate their on-board detector errors \cite{Wang2017}.

\section{Navigation}\label{sec:navigation} One of the key challenges for the APIS swarm lies in navigation, i.e., in ensuring accurate position, time and orientation estimation and control. A single satellite in the APIS mission has the visibility of GNSS satellites only for a few hours per orbit. In the absence of GNSS, all the satellites rely on their respective on-board Attitude Determination and Control Systems (ADCS), and the two-way communication with the other satellites in the swarms using Inter-satellite Links (ISLs). In this section, we discuss the clocks and time synchronization, the ADCS system on-board the satellite, and localization strategies with and without GNSS. All the data processing pertaining to the ADCS system will be done by the OBC (\Cref{sec:OBC}). In addition to the space and time estimation of each satellite, the propulsion system on-board each satellite used for orbital correction is also presented.

\subsection{Clocks and Time synchronization} The on-board clocks of the satellite swarm in the APIS mission will need to be synchronized for navigation, communication, and addressing the science mission \cite{Sundararaman2005}. The radio tomography requires accurate transmission and reception at frequencies of $10$ MHz and $30$ MHz, which is identical to the wavelengths used in interferometry in ultra-long wavelength radio astronomy \cite{Rajan2013b}. The intrinsic stochastic noise on the clock is typically measured in Allan deviation over a coherence time period \cite{Allan1966}. The Allan deviation requirements for clocks at these wavelengths are typically in the range of $10^{10}$--$10^{13}$, which is typically achieved by oven-controlled crystal oscillators or Rubidium standard clocks \cite{Allan1966}. To ensure programmable output frequencies, we propose the use of VCXO Si570, which additionally offers a low-jitter clock output for the range of $0.01$–$1.4$ GHz \cite{SiliconLabs2018}. The in-depth study of available clocks is beyond the scope of this project, and is excluded for this reason.

A solution is to align all the on-board clocks within the swarm, is clock synchronization based on time-stamping. Given a time-varying mobile network of satellite swarms, the first-order clock errors, i.e. clock offset and clock drift, can be estimated jointly along with the time-varying distances between the satellites. The satellites will employ two-way ranging to collect the transmitted and received time-stamps. Typically, these measurements are input parameters to optimization algorithms for estimating the unknown clock and distance parameters. In \cite{Rajan2015}, a constrained least squares algorithm is proposed to estimate these unknown parameters using only time-stamp measurements between the satellites. Furthermore, such algorithms can achieve time synchronization within the network, as long as each satellite has at least one communication link with any other satellite in the network. In addition, a reference for the clock is chosen arbitrarily in the network, and alternatively data-driven references are chosen from the network \cite{Rajan2015}. The achievable timing accuracy using these algorithms is directly proportional to the bandwidth of communication (i.e. number of time-stamps exchanged), and the SNR of the signals.

\subsection{Attitude Determination and Control System} \label{sec:ADCS} The Attitude Determination and Control System (ADCS) in the APIS satellites will comprise of various sensors and actuators to estimate and control the satellite attitude \cite{Kaufmann2016}. The chosen components in the ADCS system are COTS components, and have validated flight heritage.

\begin{itemize}
    \item \textit{Reaction wheels}: There are various micro-reaction wheels available on the market, which provide a small torque change, and create fine rotations, weighing $<300$ g \cite{Nudehi2008}. A favourable choice is the RWP015 from Blue Canyon Tech, which weighs as low as 130 grams and has a design life of more than five years, making it suitable for the APIS mission \cite{BC_Reactorwheels}. The APIS  mission will use the 4-wheel tetrahedron configuration such that in the event of wheel failure the mission can still continue without interruption \cite{Kaufmann2016}.

    \item \textit{Magnetic torquers}: In addition to the reaction wheels, magnetic torquers are used for attitude control, detumbling, and stabilization using the Earth’s magnetic field. The torquers consist of electromagnetic circuits without moving parts, making magnetic torquers reliable and resilient to radiation effects in comparison to other devices with sensitive electronics \cite{Inamori2012}. For the APIS mission, we chose the ISISpace iMTQ Board, which is a PCB-based 3-axis magnetic actuation and control system for 12U CubeSats \cite{ISISpace_MT}. It weighs less than $196$ grams and  is designed as a standalone detumbling system and can also be used with more advanced ADCS hardware, providing actuation of $0.2$ Am$^2$, with a magenetometer accuracy of $< 3~\mu$T.

    \item \textit{Star Trackers}: Star trackers are one of the most accurate sensors for satellite attitude estimation, as required during payload operation, Sun pointing, and for ground link communications. For the APIS mission, we chose a single NST component from Blue Canyon Tech, which offers an attitude resolution of $6$ arcseconds (cross boresight) with a Field of View (FOV) of $10^\circ \times 12^\circ$, and weighs as low as $350$ grams \cite{BC_startracker}.

    \item \textit{Sun Sensors}: In addition to Star trackers, APIS satellites will have Sun sensors on-board to provide orientation information. Sun Sensors are generally less accurate in comparison to star trackers but offer a larger FOV at typically lower costs. There will be four Sun sensors placed in the four selected corners or faces of the satellite body, which is sufficient to provide satellite orientation information with respect to the Sun. The APIS mission will use the two-axis sensor, NFS-NFSS-411, which weighs less than $35$ grams, and offers an accuracy of $0.1^\circ$  with a FOV of $140^\circ$   \cite{NSS-NFSS-411_sunsensor}.

    \item \textit{Inertial Measurement Unit (IMU)}: IMUs combine accelerometers and gyroscopes, which provide acceleration and orientation information of the satellite, respectively. In particular, MEMS-based IMUs are lightweight and reliable, with longer mission life and offer a high Technology Readiness Level (TRL). There are numerous COTS MEMS-based IMUs available in the market --- e.g., the Sensonor STIM377H, which is a tactical grade 3-axis IMU comprised of $3$ accurate MEMS gyros, $3$ high stability accelerometers and 3 inclinometers \cite{IMU}. Each APIS satellite will have $4$ of these IMU units.
\end{itemize}

\subsection{Relative localization} Huff et al. \cite{huff2017}, present a method of obtaining accurate absolute and relative position estimates of a swarm of small unmanned aerial systems. In the APIS mission, each satellite will have full access to GNSS signals for only a few hours per orbit. GNSS for navigation will therefore be unsuitable for the extended mission duration. The swarm enables collective navigation –- i.e., the satellites’ relative position to each other can be constantly determined. Each APIS satellite will require a GNSS unit, an IMU, and an ISL. When available, GNSS signal measurements will be integrated with the estimated relative positions to determine the absolute position of the satellites. The APIS satellites will use a NovAtel OEM719 multi-frequency GNSS receiver, and the on-board firmware can be reconfigured to offer sub-metre to centimetre positioning, meeting the APIS science requirements \cite{GNSS}.

For an immobile network, in the absence of a reference such as GNSS, the relative position of the satellites can be estimated using multi-dimensional scaling like algorithms \cite{Borg2013}. Furthermore, \cite{Lee2018} proposed a distributed relative position algorithm, which offers a solution to solve for the relative position of satellites cooperatively on a sphere domain. When the satellites are mobile, the relative kinematics (e.g., relative velocity and acceleration) need to be estimated by solving relative kinematics models \cite{Rajan2019}. A large consideration for the APIS team is distributed control for the swarm in terms of position and attitude. Path-planning feedback control for autonomous and distributed position control of satellite swarms will be implemented, using local sensor data to coordinate individual satellite tasks, with the assumption that each satellite is able to locally process 3D attitude and inter-satellite distances from the on-board sensors. Each satellite can evaluate, in real-time, the final target position based on the available sensor information, and safely navigate to the chosen position while avoiding collision with another satellite. This method of control uses low computational resources and autonomous position selection with safe acquisition  \cite{levchenko2018explore}, which suit the APIS mission.


\subsection{Propulsion system} \label{sec:Prop} The propulsion system consists of thrusters mainly used for orbital corrections. The satellites in the APIS mission require in-orbit corrections, and therefore, an on-board propulsion system. Two popular solutions include the Pulsed Plasma Thruster (PPT) and the Micro Electrospray Propulsion (MEP). MEPs use the principle of electrostatic extraction and acceleration of ions. The propulsion system does not require gas-phase ionization, which is an advantage. The propellant is not pressurized since it flows through capillary action. The emission is controlled by modulating the voltage applied, which provides better safety in handling the spacecraft, and light weight spacecraft \cite{NASA2019}. MEPs offer high system-level and power level efficiencies. However, there are severe system scalability issues, thus, making MEPs harder to use.

In constrast, PPTs are known for having a high specific impulse, consisting of low power electric thrusters for precise spacecraft control, and can be used for orbit maintenance \cite{Elwood2018}. The thruster uses electricity to vaporize the solid Teflon propellant to generate thrust. PPTs do not need a tank or feed system, are compact, low-cost, and consume less power in contrast to MEPs. The drawback of a PPT operation is that it may create disturbances in payload science measurements because of the possible mixing of propulsion plasma with magnetotail plasma. We can mitigate for this risk with strategic scheduling of propulsion use \cite{Spence2013}. The APIS satellites will, therefore, use a variant of PPT called Filament Pulsed Plasma Thruster (FPPT), which uses a Polytetrafluoroethylene propellant \cite{Woodruff}. Two propulsion systems will be placed aligned to the center of mass such that the system can be used for both linear and rotational movement. The thruster can also be used as a desaturating unit for the reaction wheels through the on-board computer (OBC).


\begin{table*}[tb]
\scriptsize
\centering
\caption{Comparison of various Antennas}
\label{tab19:Antenna}
\resizebox{\textwidth}{!}{%
\begin{tabular}{|l||c|c|c|c|c|}
\hline
\textbf{Manufacturers} &
  \textbf{NanoAvionics}&
  \textbf{Endurosat}&
  \textbf{Anywaves}&
  \textbf{AAC-Clyde Space}&
  \textbf{Surrey}\Tstrut\\ \hline\hline
\textbf{Website} &
  n-avionecs.com &
  endurosat.com &
  anywaves.eu &
  aac-clyde.space &
  surreysatellite.com \Tstrut\\ \hline
\textbf{\begin{tabular}[c]{@{}l@{}}Operating \\ Frequency (MHz)\end{tabular}} &
  2400-2450 &
  2400-2450 &
  2025-2290 &
  2200-2300 &
  2000-250 \Tstrut\\ \hline
\textbf{Gain (dBi)} &
  6 &
  8.3 &
  6.5 &
  7 &
  3 \Tstrut\\ \hline
\textbf{\begin{tabular}[c]{@{}l@{}}Circularly \\ Polarized\end{tabular}} &
  Yes &
  Yes &
  Yes &
  Yes &
  Yes \Tstrut\\ \hline
\textbf{Mass (g)} &
  49 &
  64 &
  123 &
  50 &
  80 \Tstrut\\ \hline
\textbf{\begin{tabular}[c]{@{}l@{}}Dimension \\ (l$\times$w$\times$h mm)\end{tabular}} & 70 $\times$ 70 $\times$ 12 & 98 $\times$ 98 $\times$ 12 & 79.8 $\times$ 79.8 $\times$ 12.1 & 81.5 $\times$ 89 $\times$ 4.1 & 82 $\times$ 82 $\times$ 20 \Tstrut\\ \hline
\end{tabular}%
}
\end{table*}

\begin{table*}[tb]
\scriptsize
\centering
\caption{Comparison of various Transceivers}
\label{tab20:Trans}
\resizebox{\textwidth}{!}{%
\begin{tabular}{|l||c|c|c|c|}
\hline
\textbf{Manufacturer} &
  \textbf{NanoAvionics}&
  \textbf{ECM} &
  \textbf{Spacecom} &
  \textbf{Skylabs}  \Tstrut\\ \hline\hline
\textbf{Website} &
  n-avionecs.com &
  ecm-space.de &
  iq-spacecom.com &
  skylabs.si \Tstrut\\ \hline
\textbf{\begin{tabular}[c]{@{}l@{}}Tx. Freq. (MHz)\end{tabular}} &
  2200-2290 &
  2200-2290 &
  2200-2290 &
  2200-2300 \Tstrut\\ \hline
\textbf{\begin{tabular}[c]{@{}l@{}}Rx. Freq. (MHz)\end{tabular}} &
  2025-2110 &
  2025-2110 &
  2025-2110 &
  2000-2100 \Tstrut\\ \hline
\textbf{Tx. Bit Rate} &
  128-512 kbps &
  20 Mbps &
  20 Mbps &
  4 Mbps  \Tstrut\\ \hline
\textbf{Modulation} &
  GMSK &
  \begin{tabular}[c]{@{}c@{}}QPSK, BPSK\end{tabular} &
  BPSK, QPSK, 8PSK &
  OQPSK \Tstrut\\ \hline
\textbf{\begin{tabular}[c]{@{}l@{}}Output Power (dBm)\end{tabular}} &
  30 &
  27 &
  30 &
  30 \Tstrut\\ \hline
\textbf{\begin{tabular}[c]{@{}l@{}}Power Consumption (W)\end{tabular}} &
  5 &
  12 &
  13 &
  5 \Tstrut\\ \cline{1-5}
\textbf{Mass (g)} &
  190 &
  190 &
  190 &
  90 \Tstrut\\ \cline{1-5}
\textbf{\begin{tabular}[c]{@{}l@{}}Dimension  \\ (l$\times$w$\times$h mm)\end{tabular}} &
  87 $\times$ 93 $\times$ 17 &
  50 $\times$ 55 $\times$ 94 &
  50 $\times$ 55 $\times$ 94 &
  95 $\times$ 91 $\times$ 10 \Tstrut\\ \cline{1-5}
\end{tabular}%
}
\end{table*}

\begin{figure}
    \centering
    \includegraphics[width=0.6\textwidth, height=0.2\textwidth]{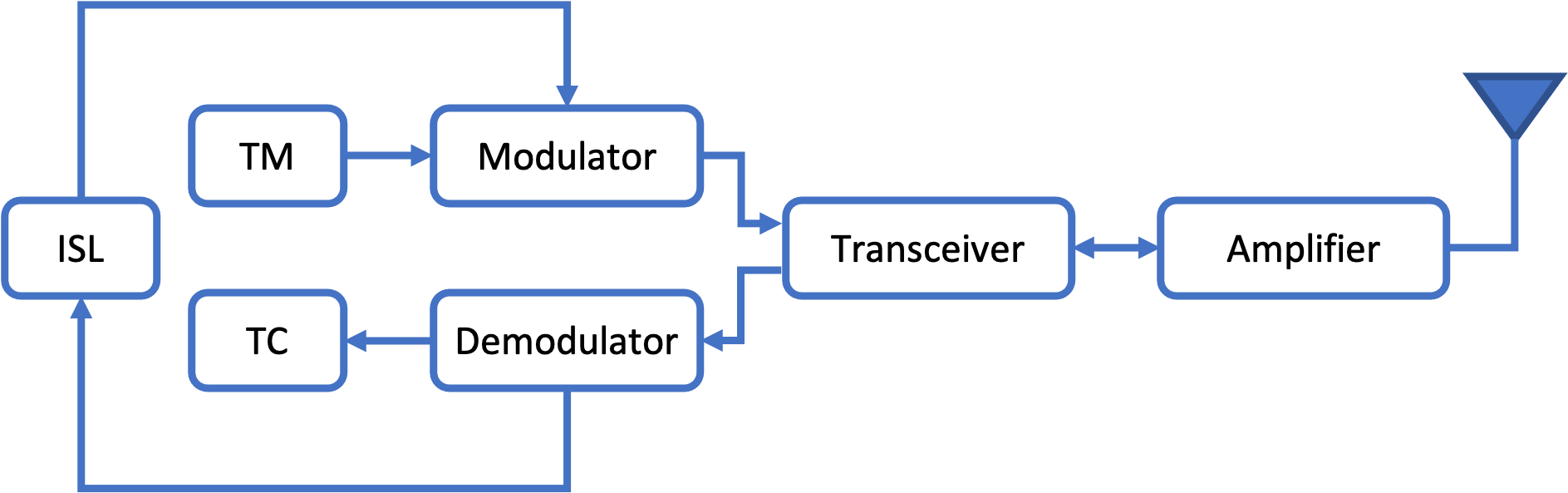}
    \caption{ \textbf{ISL communications:} The block diagram illustrates the inter-satellite link communication system, including  the S-band patch antenna, the amplifier, the modulator and demodulator blocks, and the Telemetery (TM) and Telecommand (TC) blocks, which are discussed in \Cref{sec:communication}}.
    \label{fig26:Blk}
\end{figure}

\begin{table*}
\scriptsize
    \centering
        \caption{Communication and operational requirements}
        \label{tab:communication_requirements}
        \small{
    \begin{tabular}{|m{6.5cm}|m{6cm}|}
        \hline
         \textbf{Ground station passes} & 1 pass every 2 days; 1 hour window \Tstrut \\ \hline
         \textbf{Antenna   size}  & 98 $\times$ 98 mm  \Tstrut \\ \hline
         \textbf{Data volume per pass per satellite} & 40–400 MB                            \Tstrut \\ \hline
        \textbf{Uplink and downlink frequency} & 2.45 GHz   \Tstrut \\ \hline
         \textbf{Power available for communication} & 4 W (average) \Tstrut \\ \hline
         \textbf{Downlink data rate}  & $\geq$ 20 mbit/sec \Tstrut \\ \hline
    \end{tabular}}
\end{table*}

\section{Communication} \label{sec:communication} Communication is a crucial aspect of the APIS mission for autonomy, satellite navigation, and science payload processing. A summary of all the communication and operational requirements is listed in \Cref{tab:communication_requirements}. We discuss the Inter-satellite and Earth-based communication in this section.

\subsection{Inter-satellite link (ISL)} \label{sec:ISL} Individual satellites need inter-satellite links (ISLs) to share information with each other and to transmit the information collected by scientific exploration to the ground center. Thus, the satellites also utilize satellite-to-ground communication. A simple block diagram in \Cref{fig26:Blk} shows the communication system loop for the ISL in the satellite swarm. For the scientific goals of the APIS mission, every satellite must establish a high data rate radio connection with the other satellites, and include a transceiver. The antenna must have sufficient gain and transmission power, and the transceiver must meet the data requirements. According to the existing patch antennas and transceivers available on the market, we investigated five possible antenna systems for the APIS satellites, which are listed in \Cref{tab19:Antenna}.  Considering the size of the satellite and the data transmission requirements, the S-band patch antenna produced by Endurosat is the chosen antenna for the APIS mission. Along similar lines, we considered four types of transceivers for the APIS mission as shown in \Cref{tab20:Trans}. The transceiver made by ECM Space Technologies GmbH company is attractive in terms of power, volume and mass, which is, therefore, the chosen transceiver for the APIS mission. The ECM transceiver will employ QPSK for transmission and BPSK reception, and offers a data transmission rate of up to $20$ Mbps. A summary of ISL budget for a satellite pair is presented in \Cref{tab13_4:FullLinkBudget}.

\subsection{Uplink and Downlink to Earth} \label{sec:Earth_downlink} APIS mission satellites are small in size and restricted in power supply, so at higher orbits the satellites will not be able to communicate directly with the ground-based antennas in full bandwidth. At these orbits, Mobile Ad-Hoc Network (MANET) technology can be used to send information to ground stations. The swarm uses ISLs to hop data from satellite to satellite until the satellite closest to the Earth can send the data to the ground \cite{freimann2013analysis}. During Earth fly-by at the perigee, the satellites employ multi-hop communication for data downlink, as illustrated in \Cref{fig12:CommswGrnd}. The MANET system is highly dynamic, fault-resistant, and autonomous, which is advantageous for APIS satellites to communicate data to ground stations. To reduce design complexity, the communications from satellite-to-Earth and ISLs will be combined. Delay Tolerant Networking (DTN) technology will store and transmit data back to ground stations, following processing, once satellite have reached a low-altitude location, providing reliable communication in case of network connection interruption \cite{caini2011delay}.

Using the THEMIS mission as a reference, we estimate a data rate of up to $500$ bytes/second from all of the science payloads \cite{Angelopoulos2008}. The APIS satellites can have up to 20 Mbit/s of data downlink. To manage thermal issues and data packet loss we assume a guaranteed data rate of 1 Mbit/s. With one hour of contact to the ground station, each satellite can transmit up to 400 MB. Assuming a 50\% success rate of package transmission, as commonly used \cite{Angelopoulos2008}, the data transmission per perigee pass reduces to 200 MB. Furthermore, assuming half the data consists of telemetry and housekeeping data, we can then assume a successfull transmission of up to 100 MB of science data per perigee pass. Refer to \Cref{tab13_4:FullLinkBudget} for a detailed link budget.

\begin{landscape}
\begin{table}[]
\centering
\caption{Communication Link Budgets of the satellites in the APIS mission}
\label{tab13_4:FullLinkBudget}
\resizebox{1.5\textwidth}{!}{%
\begin{tabular}{|l|c|l|l|c|l|l|c|}
\cline{1-2} \cline{4-5} \cline{7-8} \noalign{\vskip\arrayrulewidth \vskip\doublerulesep} \cline{1-2} \cline{4-5} \cline{7-8}
\multicolumn{2}{|c|}{\textbf{(a) Inter-Satellite Link}}                                               &  & \multicolumn{2}{c|}{\textbf{(b) Uplink}}                                                    &  & \multicolumn{2}{c|}{\textbf{(c) Downlink}}                                                             \Tstrut \\ \cline{1-2} \cline{4-5} \cline{7-8} \noalign{\vskip\arrayrulewidth \vskip\doublerulesep} \cline{1-2} \cline{4-5} \cline{7-8}
\textbf{ISL Frequency:}                               & \textbf{2150 MHz}                          &  & \textbf{Uplink Frequency:}                        & \textbf{2150 MHz}                    &  & \textbf{Downlink Frequency:}                          & \textbf{2150 MHz}                           \Tstrut \\ \cline{1-2} \cline{4-5} \cline{7-8} 
\textbf{Distance:}                                    & \textbf{35040.5 km}                        &  & \textbf{Distance:}                                & \textbf{35040.5 km}                  &  & \textbf{Distance:}                                    & \textbf{18846.8 km}                         \Tstrut \\ \cline{1-2} \cline{4-5} \cline{7-8} \noalign{\vskip\arrayrulewidth \vskip\doublerulesep} \cline{1-2} \cline{4-5} \cline{7-8}
\multicolumn{2}{|c|}{\textbf{1st Spacecraft:}}                                                     &  & \multicolumn{2}{c|}{\textbf{Ground Station:}}                                            &  & \multicolumn{2}{c|}{\textbf{Spacecraft:}}                                                           \Tstrut \\ \cline{1-2} \cline{4-5} \cline{7-8} \noalign{\vskip\arrayrulewidth \vskip\doublerulesep} \cline{1-2} \cline{4-5} \cline{7-8}
\multirow{3}{*}{Spacecraft Transmitter Power Output:} & 8.0 W                                      &  & \multirow{3}{*}{Transmitter Power Output:}        & 50.0 W                               &  & \multirow{3}{*}{Spacecraft Transmitter Power Output:} & 8.0 W                                       \Tstrut \\ \cline{2-2} \cline{5-5} \cline{8-8} 
                                                      & 9.0 dBW                                    &  &                                                   & 17.0 dBW                             &  &                                                       & 9.0 dBW                                     \\ \cline{2-2} \cline{5-5} \cline{8-8} 
                                                      & 39.0 dBm                                   &  &                                                   & 47.0 dBm                             &  &                                                       & 39.0 dBm                                    \\ \cline{1-2} \cline{4-5} \cline{7-8} 
Spacecraft Transmission Line Losses:                  & -1.0 dB                                    &  & Transmission Line Losses:                         & -3.0 dB                              &  & Spacecraft Transmission Line Losses:                  & -1.0 dB                                     \Tstrut \\ \cline{1-2} \cline{4-5} \cline{7-8} 
S/C Connector, Filter or In-Line Switch Losses:       & 0.0 dB                                     &  & S/C Connector, Filter or In-Line Switch Losses:   & -1.0 dB                              &  & S/C Connector, Filter or In-Line Switch Losses:       & 0.0 dB                                      \Tstrut \\ \cline{1-2} \cline{4-5} \cline{7-8} 
Spacecraft Antenna Gain:                              & 8.3 dBiC                                   &  & Antenna Gain:                                     & 31.4 dBiC                            &  & Spacecraft Antenna Gain:                              & 8.3 dBiC                                    \Tstrut \\ \cline{1-2} \cline{4-5} \cline{7-8} 
Spacecraft EIRP:                                      & 46.3 dBm                                   &  & Ground Station EIRP:                              & 74.3 dBW                             &  & Spacecraft EIRP:                                      & 46.3 dBm                                    \Tstrut \\ \cline{1-2} \cline{4-5} \cline{7-8} \noalign{\vskip\arrayrulewidth \vskip\doublerulesep} \cline{1-2} \cline{4-5} \cline{7-8}
\multicolumn{2}{|c|}{\textbf{Crosslink Path:}}                                                     &  & \multicolumn{2}{c|}{\textbf{Uplink Path:}}                                               &  & \multicolumn{2}{c|}{\textbf{Downlink Path:}}                                                        \Tstrut \\ \cline{1-2} \cline{4-5} \cline{7-8} \noalign{\vskip\arrayrulewidth \vskip\doublerulesep} \cline{1-2} \cline{4-5} \cline{7-8}
Spacecraft Antenna Pointing Loss:                     & -1.0 dB                                    &  & Ground Station Antenna Pointing Loss:             & -1.0 dB                              &  & Spacecraft Antenna Pointing Loss:                     & -1.0 dB                                     \Tstrut \\ \cline{1-2} \cline{4-5} \cline{7-8} 
Antenna Polarization Loss:                            & -1.5 dB                                    &  & Antenna Polarization Loss:                        & -4.0 dB                              &  & Antenna Polarization Loss:                            & -1.5 dB                                     \Tstrut \\ \cline{1-2} \cline{4-5} \cline{7-8} 
Path Loss:                                            & -190.0 dB                                  &  & Path Loss:                                        & -190.0 dB                            &  & Path Loss:                                            & -184.6 dB                                   \Tstrut \\ \cline{1-2} \cline{4-5} \cline{7-8} 
Atmospheric Loss:                                     & 0 dB                                       &  & Atmospheric Losses:                               & -3 dB                                &  & Atmospheric Loss:                                     & -2.2 dB                                     \Tstrut \\ \cline{1-2} \cline{4-5} \cline{7-8} 
Isotropic Signal Level at Spacecraft:                 & -146.4 dBm                                 &  & Isotropic Signal Level at Ground Station:         & -124.7 dBm                           &  & Isotropic Signal Level at Ground Station:             & -143.2 dBm                                  \Tstrut \\ \cline{1-2} \cline{4-5} \cline{7-8} \noalign{\vskip\arrayrulewidth \vskip\doublerulesep} \cline{1-2} \cline{4-5} \cline{7-8}
\multicolumn{2}{|c|}{\textbf{2nd Spacecraft:}}                                                     &  & \multicolumn{2}{c|}{\textbf{Spacecraft:}}                                                &  & \multicolumn{2}{c|}{\textbf{Ground Station:}}                                                       \Tstrut \\ \cline{1-2} \cline{4-5} \cline{7-8} \noalign{\vskip\arrayrulewidth \vskip\doublerulesep} \cline{1-2} \cline{4-5} \cline{7-8}
2nd Spacecraft Antenna Pointing Loss:                 & 0.0 dB                                     &  & Spacecraft Antenna Pointing Loss:                 & 0.0 dB                               &  & Ground Station Antenna Pointing Loss:                 & -2.0 dB                                     \Tstrut \\ \cline{1-2} \cline{4-5} \cline{7-8} 
2nd Spacecraft Antenna Gain:                          & 8.3 dBiC                                   &  & Spacecraft Antenna Gain:                          & 8.3 dBiC                             &  & Ground Station Antenna Gain:                          & 31.35 dBic                                  \Tstrut \\ \cline{1-2} \cline{4-5} \cline{7-8} 
2nd Spacecraft Transmission Line Losses:              & -1 dB                                      &  & Spacecraft Transmission Line Losses:              & -1 dB                                &  & Ground Station Transmission Line Losses:              & -1 dB                                       \Tstrut \\ \cline{1-2} \cline{4-5} \cline{7-8} 
S/C Transmission Line Coefficient:                    & 0.7943                                     &  & S/C Transmission Line Coefficient:                & 0.7943                               &  & G.S. Transmission Line Coefficient:                   & 0.7943                                      \Tstrut \\ \cline{1-2} \cline{4-5} \cline{7-8} 
2nd Spacecraft Effective Noise Temperature:           & 250 K                                      &  & Spacecraft Effective Noise Temperature:           & 250 K                                &  & Ground Station Effective Noise Temperature:           & 542 K                                       \Tstrut \\ \cline{1-2} \cline{4-5} \cline{7-8} 
2nd Spacecraft Figure of Merit (G/T):                 & -16.7 dB/K                                 &  & Spacecraft Figure of Merit (G/T):                 & -16.7 dB/K                           &  & Ground Station Figure of Merit (G/T):                 & 3.0 dB/K                                    \Tstrut \\ \cline{1-2} \cline{4-5} \cline{7-8} 
S/C Signal-to-Noise Power Density (S/No):             & 65.5 dBHz                                  &  & S/C Signal-to-Noise Power Density (S/No):         & 87.3 dBHz                            &  & G.S. Signal-to-Noise Power Density (S/No):            & 86.4 dBHz                                   \Tstrut \\ \cline{1-2} \cline{4-5} \cline{7-8} 
\multirow{2}{*}{System Desired Data Rate:}            & $1.00 \times 10^7$ bps                     &  & \multirow{2}{*}{System Desired Data Rate:}        & $2.00 \times 10^7$ bps               &  & \multirow{2}{*}{System Desired Data Rate:}            & $2.00 \times 10^7$ bps                      \Tstrut \\ \cline{2-2} \cline{5-5} \cline{8-8} 
                                                      & 50.0 dBHz                                  &  &                                                   & 73.0 dBHz                            &  &                                                       & 73.0 dBHz                                   \Tstrut \\ \cline{1-2} \cline{4-5} \cline{7-8} 
System Required Eb/No:                                & 10 dB                                      &  & Telemetry System Required Eb/No:                  & 10 dB                                &  & Telemetry System Required Eb/No:                      & 10 dB                                       \Tstrut \\ \cline{1-2} \cline{4-5} \cline{7-8} \noalign{\vskip\arrayrulewidth \vskip\doublerulesep} \cline{1-2} \cline{4-5} \cline{7-8}
\multicolumn{1}{|r|}{\textbf{System Link Margin:}}    & \multicolumn{1}{l|}{\textbf{5.5 dB}} &  & \multicolumn{1}{r|}{\textbf{System Link Margin:}} & \multicolumn{1}{l|}{\textbf{4.2 dB}} &  & \multicolumn{1}{r|}{\textbf{System Link Margin:}}     & \multicolumn{1}{l|}{\textbf{3.4 dB}} \Tstrut \\ \cline{1-2} \cline{4-5} \cline{7-8} \noalign{\vskip\arrayrulewidth \vskip\doublerulesep} \cline{1-2} \cline{4-5} \cline{7-8}
\end{tabular}%
}
\end{table}
\end{landscape}

\section{On-board processing} \label{sec:OBC} The On-Board Computer (OBC) is a crucial component to the functionality of the mission and is responsible for reading, collecting, and processing sensor information from the ADCS, GNSS, and communications systems, which includes executing Telemetry, Tracking and Command (TTC) operations, and storing on-board health data. Additionally, the OBC is responsible for maintaining the On-Board Timers (OBT) for synchronization, localization, control, and performing inertial referencing calculations \cite{Burlyaev2012}. In addition, the OBC must execute all expected tasks while abiding to bus requirements. Thus, we need to keep the power consumption and mass low, with the frequency and Total Ionizing Dose (TID) high. Higher frequency ensures tasks can be completed within the lowest necessary time, however, this requires high power consumption. Thus, both frequency and power must be balanced.

Since the APIS mission is to study the magnetosphere outside of LEO, radiation hardening of the OBC processor is an important requirement, as the charged particles ejected from the Sun can cause detrimental and undesired effects to the on-board electronics \cite{Keys2008,Botma2011a}. Based on the NASA RHESE project, radiation hardening of OBC components is recommended through: material hardening of OBC components and shielding; the inclusion of redundant hardware; software verification and reconfigurability \cite{Keys2008}. The measure of accumulated radiation a device can withstand prior to becoming unreliable is known as the TID, which acts as a metric for the life expectancy of the device \cite{Botma2011a}, and sets the parameters for radiation hardening. A higher TID ensures a higher life expectancy for the electronics.

A trade study was completed to compare the available options viable for the APIS mission listed in \Cref{tab17:OBCSpec}. All three options have similar dimensions, though, ISISpace offers the highest operating frequency at the cost of higher power consumption. However, for the APIS mission, mass is a critical criteria, and hence we chose the IMT component, weighing nearly half the mass of the remaining two options.

The satellite swarm in the APIS mission will comprise of autonomous satellites, and to that end will each achieve (a) Self-configuration, (b) Self-optimization, (c) Self-healing and (d) Self-protection, during mission operation \cite{Truszkowski2004a}. If an autonomous system has the authority to make changes to its command, and behaves in an unexpected way, it will reduce the level of trust between the system and its operators \cite{alves2018considerations}. Changes in trust levels will complicate mission design and may necessitate the creation and installation of safeguards that further increase system complexity. In addition to exploiting autonomy, the APIS mission will facilitate distributed functionality, enabling emergent behavior of the satellite swarm.

To enable the emergent behavior of the swarm satellites, we need to partner an appropriate operating system with the OBC. NASA conducted research on the feasibility of using COTS technology for the Command and Data Handling (CDH) system of a swarm of satellites \cite{Cockrell2012a}. The study used an android operating system with specific programming to relay data from both satellite-to-satellite and satellite-to-ground stations, demonstrating the feasibility of the satellite swarm behavior using COTS software. We recommend using an adapted version of FreeRTOS, which can process, allocate, and communicate tasks in a large-scale multi-processor network with reduced network congestion and communication energy \cite{Babuscia2015}.

\begin{table*}[tb]
\scriptsize
    \centering
    \caption{Potential OBC options and specifications}
    \label{tab17:OBCSpec}
    \small{
    \begin{tabular}{|p{3cm}|c|c|c|c|c|}
    \hline
        \textbf{OBC Option} &
          \textbf{\begin{tabular}[c]{@{}c@{}}Power \\ Consumption \\ (mW)\end{tabular}} &
          \textbf{Mass (g)} &
          \textbf{\begin{tabular}[c]{@{}c@{}}Dimensions\\ $\mathbf{l \times w \times h}$ \\ (mm) \end{tabular}} &
          \textbf{\begin{tabular}[c]{@{}c@{}}Frequency \\ (MHz)\end{tabular}} &
          \textbf{\begin{tabular}[c]{@{}c@{}}Total \\ ~Ionizing~ \\ Dose \\ (kRad)\end{tabular}} \\ \hline\hline
        \textbf{ISIS}         & 400 & 76    & $96 \times 90 \times 12.4$ & 400 & N/A \Tstrut\\ \hline
        \textbf{CubeComputer} & 200 & 50-70 & $90 \times 96 \times 10$   & 48  & 20  \Tstrut\\ \hline
        \textbf{IMT}          & 300 & 38    & $96 \times 90 \times 10$   & 200 & 15  \Tstrut\\ \hline
    \end{tabular}
    }
\end{table*}


\begin{table}
\scriptsize
\centering
\caption{Power budget}
\label{tab:PowerBudg}
    \begin{tabular}{|l | c | c | c|}
    \hline
        \textbf{System} & \textbf{Average (W)} & \textbf{Idle (W)} & \textbf{Peak (W)} \TBstrut \\
        \hline
        \hline
        Payload & 10 & 1 & 250  \Tstrut \\
        DHS &	2   &	2 &	2 \\
        IMU &	2.4	&   2.4 &  	2.4 \\
        RF/ISL Link &	13 &	1.7 & 38 \\
        ADCS/OBC & 10 &	10 & 10 \\
        Reaction Wheels	& 2.4 &	2.4 &	4 \\
        Magnetic Torquers &	0.175 &	0.175 &	1.2 \\
        Thermal & 8 & 5 & 15 \\
        Power Electronics & 4 & 4 & 4 \\
        Battery Charge & 10 & 5 & 20 \\
        2\% Losses & 1.2395 & 0.6735 & 7.892 \Bstrut \\
        \hline\hline
        \textbf{Total Load:} & \textbf{63.2} & \textbf{34.3} & \textbf{354.4} \TBstrut \\
        \hline
    \end{tabular}
\end{table}

\section{Power system} \label{sec:power_system}
The various sub-systems in the APIS satellite, e.g., navigation, communication and OBC are fueled by the on-board power system. The electrical power system deals with the generation, storage, monitoring, control, and distribution of electrical power. A summary of the power budget requirements for various APIS satellite sub-systems and all the scientific payloads discussed in the previous sections, are listed in \Cref{tab:PowerBudg}. The total average power demand for an APIS satellite is estimated at 63.2 watts. We briefly introduce various sub-systems within the power system.

\begin{figure}[]
\centering
\begin{subfigure}{.5\textwidth}
  \centering
  \includegraphics[width=.8\linewidth]{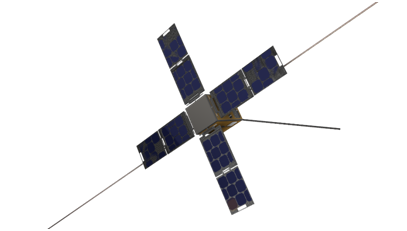}
  \caption{}
  \label{fig20:SolPanCAD}
\end{subfigure}%
\begin{subfigure}{.5\textwidth}
  \centering
  \includegraphics[width=.8\linewidth]{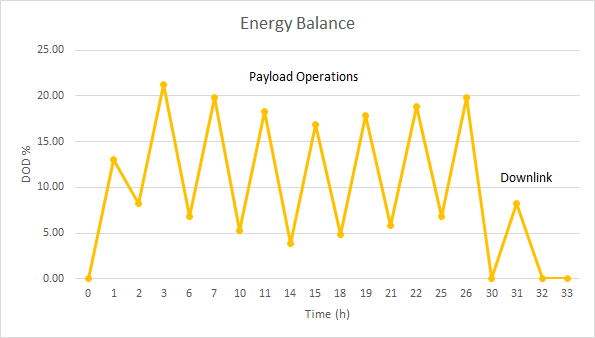}
  \caption{}
  \label{fig:sub2}
\end{subfigure}
\caption{ \textbf{(a) APIS satellite solar panels:} An illustration of the solar panels deployed in an APIS satellite. The solar panels are fixed in a windmill fashion to ensure the panels are consistently facing the Sun during orbit, gaining maximal solar energy for powering satellite system. \textbf{(b) APIS satellite energy balance:} The typical energy balance for an APIS satellite battery over the course of one orbit is presented.}
\label{fig25:BattEngy}
\end{figure}



\begin{itemize}
    \item \textit{Generation}: Solar panels are the safest, and cheapest choice for the missions for near-Earth orbit. The state of the art AM0 ATJ (Advanced Triple Junction) or ZQJ (Azur Quad Junction) solar cells available give greater than $35$\% beginning of life efficiency, and will produce $245$ W/m$^2$ \cite{AzurSpace2019}. The battery acts as a backup support whenever the solar cells or array unable to point towards the Sun during maneuvers or payload operations. The solar panel is deployable and fixed, instead of variable, for Sun-tracking. Moreover, body-mounted solar cells cannot meet the power requirements of the satellite swarm, thus,  deployable solar panels are necessary. There are two panels for each side in a fixed configuration, and the solar panel parameters are presented in \Cref{tab15:SolPanParams}. We have fixed a windmill design of the solar panels such that they are always pointing towards the Sun during their orbit, which is illustrated in \Cref{fig20:SolPanCAD}.

    \item \textit{Storage}: Lithium Ion (Li-Ion) cells are reliable in both small and large satellite applications, and their performance can be further enhanced with carbon nanotube electrodes \cite{Xiong2013}. \Cref{fig25:BattEngy} shows the energy balance for the battery, where the estimated charge and discharge cycles are illustrated during mission life. For a five-year mission, a safe limit with a margin for Depth of Discharge (DoD) of $20$\% is about $30000$ cycles. With these constraints, we choose the LG18650MJ1 Li-Ion cell with the battery sizing as listed in \Cref{tab16:BattParams}.


    \item \textit{Power control and monitoring}: Traditionally, the power control unit regulates power from the solar panels and batteries, consisting of three units to adjust the output and charge the battery pack. A more optimal system includes a digitally integrated array control and battery charging unit \cite{Koka2016}. Additionally, the system determines the health of critical power elements and has a set of protection against overloads, undervoltage, and overvoltage occurrences.

    \item \textit{Power distribution}: Power distribution systems can be distributed or centralized. The best choice for the APIS satellites are Solid-State Power Controllers (SSPC), as they are more reliable than electromechanical relays and are re-triggerable in case of faults caused by noise, electromagnetic interference or radiation effects \cite{Glass2010}. For satellites of 12U size, a centralized, regulated distribution will be efficient in mass and volume with all bus systems powered with a single Direct Current/Direct Current (DC/DC) module. We choose dedicated DC/DC converters for redundant systems to avoid single-point failures, where one failure could otherwise affect the entire satellite. The payload is sensitive, requiring completely different voltage scales, and should include local DC/DC converters, a low drop-out regulator, and a switch to disconnect the system from the source for isolating a fault if required.
\end{itemize}

\begin{table}
\scriptsize
\parbox{.5\linewidth}{
    \centering
    \caption{Solar panel parameters}
    \label{tab15:SolPanParams}
        \begin{tabular}{|l|c|}
        \hline
            \textbf{Parameter}        & \textbf{Value}    \TBstrut \\ \hline \hline
            Face area                 & 0.04 m$^2$   \Tstrut\\ \hline
            Generation @ BOL          & 78.5 W    \Tstrut\\ \hline
            ATJ/ZQJ 0.5 mA cells in parallel & 4       \Tstrut\\ \hline
            Number of strings         & 4        \Tstrut\\ \hline
            Regulation control        & IAC-BC   \Tstrut \\ \hline
            \begin{tabular}[c]{@{}l@{}}Total Panel Area\\ (2 panels $\times$ 4 panels)\end{tabular} & 0.32 m$^2$   \\ \cline{1-2}
            Generation @ EOL          & 67 W      \Tstrut\\ \hline
            Number of cells in series & 14       \Tstrut\\ \hline
            Power loss per string     & 16 W      \Tstrut\\ \hline
            Estimated weight          & 1.4 kg    \Tstrut\\ \hline
        \end{tabular}
}
\hfill
\parbox{.5\linewidth}{
    \centering
    \caption{Battery parameters}
    \label{tab16:BattParams}
    \begin{tabular}{|l|c|}
    \hline
        \textbf{Parameter}          & \textbf{Value}     \TBstrut \\ \hline \hline
        Allowed   DoD               & \textless 20\%   \Tstrut\\ \hline
        Load   power                   & \begin{tabular}[c]{@{}c@{}} \textbf{Peak:} $\sim$54 W\\ \textbf{Idle:} 34 W\end{tabular}             \\ \hline
        Discharge at average voltage & \begin{tabular}[c]{@{}c@{}} \textbf{Peak:} 10 A (100 ms/3 s)\\ \textbf{Nominal:}  1.1 A \end{tabular} \\ \hline
        Regulation   control        & IAC-BC              \Tstrut \\ \hline
        Final   battery size        & 8S4P                \Tstrut \\ \hline
        Number   of cells in series & 8                   \Tstrut \\ \hline
        Total   cycles              & $\sim$30   000      \Tstrut \\ \hline
        Cell                        & LG18650MJ1   3.5 Ah \Tstrut \\ \hline
        Capacity                    & 400   Wh            \Tstrut \\ \hline
    \end{tabular}
}
\end{table}

\begin{figure}[tb]
    \centering
    \includegraphics[width=0.75\textwidth]{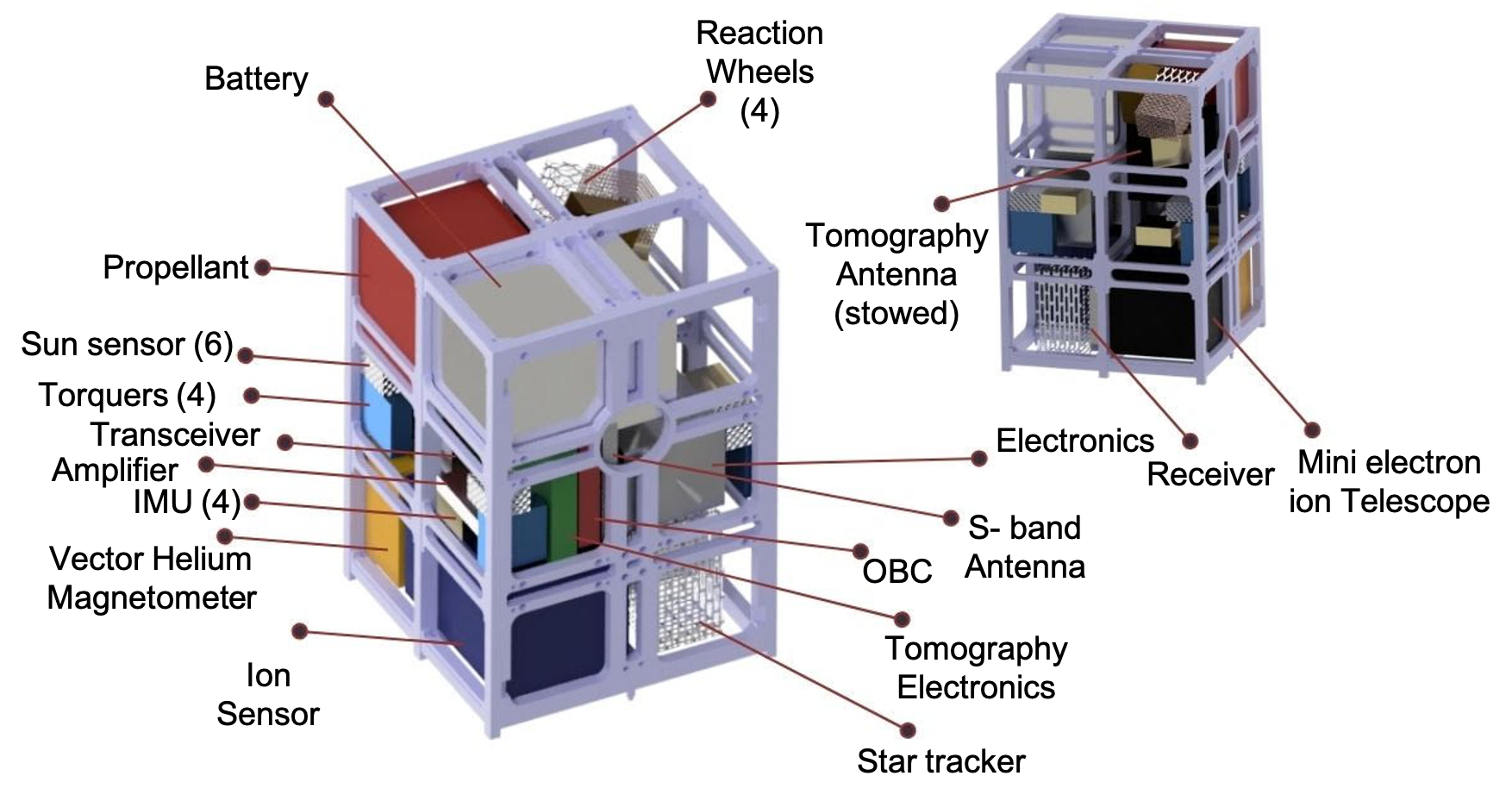}
    \caption{ \textbf{APIS satellite decomposition:} A simplified view of the APIS 12U satellite along with the subsystems and scientific payloads. A list of all chosen COTS components are listed in \Cref{tab:MassBudg}.}
    \label{fig25:satellite_12U}
\end{figure}



\begin{table}[tb]
\small
\centering
\caption{Mass budget}
\label{tab:MassBudg}
\begin{tabular}{|c|c|c|c|c|}
\hline
\multirow{2}{*}{\textbf{System}}
& \multirow{2}{*}{\textbf{Component}}
& \textbf{Mass}
& \textbf{Dimension}
& \textbf{Volume}
\\ \cline{3-5}
{}
& {}
& \textbf{(kg)}
& \textbf{(mm)}
& \textbf{(mm${^3}$)}
\\ \hline
\multirow{7}{*}{Payload} & Ion Sensor & 1.37 & 100 x 100 x 150 & 500000 \\ \cline{2-5}
 & Vector Helium Magnetometer & 0.82 & 100 x 100 x 50 & 500000 \\ \cline{2-5}
 & Mini Electron ion Telescope & 1.1 & 100 x 100 x 90 & 900000 \\ \cline{2-5}
 & Tomography Electronics & 1 & 100 x 100 x 20 & 200000 \\ \cline{2-5}
 & Tomography Antenna & \multirow{2}{*}{1} & 15000 x 27 (deployed) & - \\ \cline{2-2} \cline{4-5}
 & Tomography Antenna &  & 90 x 90 x 40 (stowed) & 324000 \\ \cline{2-5}
 & Magnetic Boom & 0.4 & 8 x 680 & 5440 \\ \hline
OBC & - & 0.5 & 96 x 90 x 17 & 146880 \\ \hline
\multirow{5}{*}{ADCS} & Reaction Wheels (4   nos) & 0.52 & 42 x 42 x 19 & 33516 \\ \cline{2-5}
 & Torquers (4 no.) & 0.8 & 96 x 90 x 17 & 146880 \\ \cline{2-5}
 & Star Tracker (1 no.) & 0.36 & 100 x 55 x 50 & 275000 \\ \cline{2-5}
 & Sun Sensor (6 nos) & 0.21 & 34 x 40 x 20 & 27200 \\ \cline{2-5}
 & IMU (4 no.) & 0.22 & 39 x 45 x 22 & 38610 \\ \hline
\multirow{2}{*}{GNSS} & Receiver & 0.1 & 71 x 46 x 11 & 35926 \\ \cline{2-5}
 & Antenna & 0.1 & - & - \\ \hline
Frame   structure & - & 1.5 & 226 x 226 x 340 & - \\ \hline
\multirow{3}{*}{Communications} & S-Band Antenna & 0.07 & 98 x 98 x 12 & 115248 \\ \cline{2-5}
 & Amplifier & 0.3 & 68 x 45 x 12 & 36720 \\ \cline{2-5}
 & Transceiver & 0.1 & 95 x 50 x 55 & 261250 \\ \hline
\multirow{4}{*}{Power   system} & \multirow{2}{*}{Solar Panel} & \multirow{2}{*}{1.4} & 200  x & 40000 \\ \cline{4-5}
 &  &  & 200  (8 nos) & - \\ \cline{2-5}
 & Battery & 1.8 & 96 x 96 x 144 & 1327104 \\ \cline{2-5}
 & Electronics & 0.5 & 80 x 80 x 70 & 448000 \\ \hline
Thermal   Control & - & 1.5 & N/A & - \\ \hline
\multirow{2}{*}{Propulsion   System} & Propellant & 1.2 & 96 x 96 x 96 & 884736 \\ \cline{2-5}
 & Dry Mass & 4 & - & - \\ \hline
Misc/Harness & - & 1 & - & - \\ \hline
 & \textbf{Total} & \textbf{21.87} & \textbf{} & \textbf{6246510} \\ \hline
\end{tabular}
\end{table}

\section{Conclusion} \label{sec:conclusion} The APIS mission aims to characterize plasma processes and measure the effect of the solar wind on Earth's magnetosphere. We propose a swarm of 40 CubeSats in two highly-elliptical orbits around the Earth, to investigate both the magnetotail and the Sun-ward magnetosphere, with a focus on radio tomography in the magnetotail. The APIS satellites will employ radio tomography, and carry various in-situ measurement systems on-board, e.g., super-thermal ion spectrograph, miniaturized electron and ion telescope, and vector helium magnetometer to provide large-scale maps of plasma density and turbulence in the magnetotail. We estimate the minimum science requirement, of radio tomography in the magnetotail to be achieved in the course of the first 3–4 months into the mission. At the end of the first year, all the key science phases will be completed. 

Towards the satellite design, various subsystems have been studied, e.g., navigation, communication, on-board processing, power and propulsion systems. A preliminary investigation on the mass and volume budget of a single satellite was conducted, and a summary of all  payloads and components are listed in \Cref{tab:MassBudg}. We show that COTS technologies available today can be used to meet the outlined science requirements.  \Cref{tab:MassBudg} indicates that the APIS mission can be designed using homogeneous CubeSats, with a possible  12U  form  factor  (226  mm $\times$ 226  mm $\times$ 340  mm). A satellite decomposition is shown in \Cref{fig25:satellite_12U}, which encompasses all the subsystems discussed in the previous sections and the scientific payloads. We favor the skin-frame structure over a simple frame structure because of its resistance to bending, tensional and axial forces \cite{Stevens2002}, and an aluminum and carbon-fiber-epoxy as the material for our bus. The APIS project is a partially funded by NASA, though this is beyond the scope of this work.


\section{Future work} A comprehensive system design is an on-going investigation, and will be presented in subsequent work. As stated earlier, the power and mass budgets are preliminary and need a through review. Furthermore, we would like to investigate various thermal control methods (including both active and passive systems), suitable for the APIS mission. In addition to a detailed system model, some of the key future challenges involve distributed algorithms for swarm behaviour, including energy-efficient communication \cite{kodheli2020satellite}, smart autonomous navigation \cite{reid2020satellite}, distributed processing \cite{krupke2019automated}, attitude control \cite{vedant2019reinforcement}, and in-depth mission planning \cite{bernhard2020coordinated,sung-hoon-2020}. Furthermore, significant advances in propulsion technology \cite{dono2018propulsion}, and miniaturization of satellite subsystems and payloads in the upcoming decades will yield smaller satellites and larger swarms, subsequently benefiting the APIS mission \cite{levchenko2018explore}.

\begin{figure}
    \centering
    \includegraphics[width=0.4\textwidth]{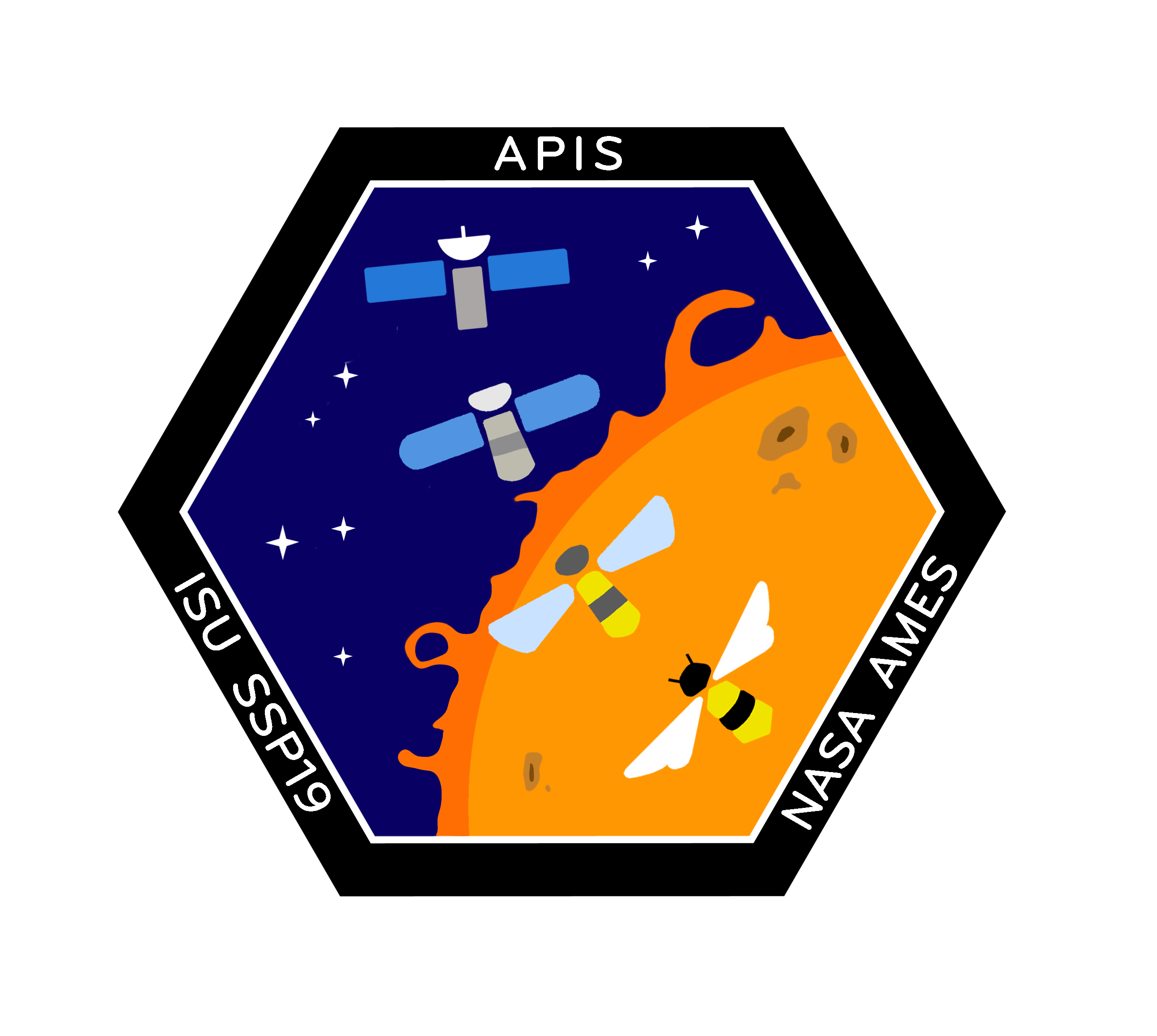}
    \caption{ \footnotesize \textbf{APIS mission patch:} APIS is adopted from the genus name \textit{Apis}, meaning `Bee' in Latin, which was an inspiration for the APIS mission patch designed during this study}
    \label{fig:APIS_patch}
\end{figure}

\section{Acknowledgement} The APIS mission design study was part of the International Space University Space Studies Program in 2019 (ISU-SSP19) as the Next Generation Space Systems: Swarms Team Project, which is partially funded by NASA. This 9-week intense project was conducted by a team of 32 members from 16 nationalities, who are drawn from diverse professional backgrounds within the space sector, which include but are not limited to engineers, scientists, lawyers and managers. The authors of this publication include the APIS team, the Chairs Dr. Jacob Cohen, Anh Nguyen and Andrew Simon-Butler, and the support staff Ruth McAvinia. The name of the project is adopted from the genus name \textit{Apis}, meaning `Bee' in Latin, which was an inspiration for the APIS mission patch designed during this study (see \Cref{fig:APIS_patch}). \par
The authors would like to thank the Editor, the anonymous reviewers, and Dr. Prem Sundaramoorthy for their constructive comments, which helped improve the quality this article. 

\bibliographystyle{elsarticle-num}
\bibliography{core/bibliography}
\end{document}